\documentclass[journal]{elsarticle}

\usepackage[utf8]{inputenc}
\usepackage{url}
\usepackage{amsmath}
\usepackage{amssymb}
\usepackage{color}
\usepackage{graphicx}
\usepackage[official]{eurosym}
\usepackage{gensymb}
\usepackage{booktabs}
\usepackage{subcaption}
\usepackage{siunitx}
\usepackage{mathtools}
\usepackage{multirow}
\usepackage{algorithm}
\usepackage{algpseudocode}
\usepackage{enumitem}
\usepackage{threeparttable}
\usepackage{physics}

\algrenewcommand\algorithmicrequire{\textbf{Input}}
\algrenewcommand\algorithmicensure{\textbf{Output}}

\DeclarePairedDelimiter\ceil{\lceil}{\rceil}

\definecolor{darkcerulean}{rgb}{0.03, 0.27, 0.49}
\definecolor{ao}{rgb}{0.0, 0.5, 0.0}
\definecolor{bred}{rgb}{0.8, 0.25, 0.33}
\definecolor{crimson}{rgb}{0.86, 0.08, 0.24}
\definecolor{mygray}{gray}{0.6}

\newcommand{\review}[1]{\textcolor{black}{#1}}

\begin{document}

\title{Assessing the Impact of Offshore Wind Siting Strategies on the Design of the European Power System}

\author[ulg]{David~Radu\corref{cor1}}
\ead{dcradu@uliege.be}
\author[ulg]{Mathias~Berger}
\author[ulg]{Antoine Dubois}
\author[ulg]{Raphaël Fonteneau}
\author[uzg]{Hrvoje Pandžić}
\author[nyu]{Yury Dvorkin}
\author[ulg]{Quentin Louveaux}
\author[ulg]{Damien Ernst}

\cortext[cor1]{Corresponding author.}
\address[ulg]{Department of Electrical Engineering and Computer Science, University of Liège, Allée de la Découverte 10, 4000 Liège, Belgium}
\address[uzg]{Faculty of Electrical Engineering and Computing, University of Zagreb, HR-1000 Zagreb, Croatia}
\address[nyu]{Department of Electrical and Computer Engineering, NYU Tandon School of Engineering, NY 11201, USA}

\begin{abstract}
This paper provides a detailed account of the impact of different offshore wind siting strategies on the design of the European power system. To this end, a two-stage method is proposed. In the first stage, a highly-granular siting problem identifies a suitable set of sites where offshore wind plants could be deployed according to a pre-specified criterion. Two siting schemes are analysed and compared within a realistic case study. These schemes essentially select a pre-specified number of sites so as to maximise their aggregate power output and their spatiotemporal complementarity, respectively. In addition, two variants of these siting schemes are provided, wherein the number of sites to be selected is specified on a country-by-country basis rather than Europe-wide. In the second stage, the subset of previously identified sites is passed to a capacity expansion planning framework that sizes the power generation, transmission and storage assets that should be deployed and operated in order to satisfy pre-specified electricity demand levels at minimum cost. Results show that the complementarity-based siting criterion leads to system designs which are up to 5\% cheaper than the ones relying the power output-based criterion when offshore wind plants are deployed with no consideration for country-based deployment targets. On the contrary, the power output-based scheme leads to system designs which are consistently 2\% cheaper than the ones leveraging the complementarity-based siting strategy when such constraints are enforced. The robustness of the results is supported by a sensitivity analysis on offshore wind capital expenditure and inter-annual weather variability, respectively.
\end{abstract}

\begin{keyword}
asset siting, offshore wind, resource complementarity, capacity expansion planning
\end{keyword}

\maketitle

\section{Introduction}

\review{The large-scale deployment of technologies harnessing renewable energy sources (RES) for electricity production has been a mainstay of climate and decarbonization policies. In Europe, solar photo-voltaic and onshore wind power plants have formed the bulk of new renewable capacity additions for over a decade \cite{IEA2020b}. Nevertheless, in spite of the need for extra capacity deployments required to achieve ambitious decarbonization targets \cite{IEA2050}, the pace at which these technologies are deployed in a number of countries has remained sluggish of late \cite{IEA2020b}, often as a result of social acceptance issues \cite{Segreto2020} and the phasing out of renewable support schemes. On the other hand, the economics of offshore wind power generation have greatly improved in recent years \cite{IRENA2020}. Offshore wind power plants are also located in unpopulated areas and are therefore less subject to social acceptance issues than onshore ones. Furthermore, the offshore wind resource is most often of much better quality than the onshore one \cite{globalwindatlas}. Hence, the large-scale deployment of offshore wind power plants has increasingly been viewed as a key enabler of European decarbonization efforts \cite{EC_2018,WindEurope2020_main}.}

\review{However, widely-available RES such as solar irradiance or offshore wind are inherently variable on time scales ranging from minutes to years and integrating them in power systems typically complicates planning and operational procedures \cite{Engeland2017}. Several solutions have been advocated to alleviate these issues, including the large-scale deployment of electricity storage systems \cite{Geth2015,Stenclik2017} or the implementation of demand response programs \cite{OConnell2014}. Alternatively, since RES are heterogeneously-distributed in space and time, it has been suggested that siting RES electricity production assets so as to exploit this diversity may reduce the aggregate output variability of RES power plants as well as the residual electricity load (i.e., total load minus renewable production) \cite{Milligan1999,Giebel2001}. The concept of RES complementarity formalises this idea \cite{Jurasz2020}.}

\review{From a modelling perspective, the interplay between investment (both siting and sizing) and operational decisions should be accounted for in order to evaluate the impact of siting strategies on system design and economics. Hence, ideally, models should perform both siting and sizing simultaneously, have a high spatiotemporal resolution as well as a high level of technical detail. Unfortunately, such models quickly become impractical (e.g., require tens of thousands of core-hours \cite{MacDonald2016}) or intractable. Thus, the siting and sizing problems have traditionally been tackled separately in the literature, but the outcomes of siting models have rarely been leveraged in sizing models.}

\review{In this paper, the role that offshore wind power plants may play in the European power system is analysed, with a particular focus on the impact that plant siting strategies have on system design and economics. To this end, a two-stage method is developed. In the first stage, a highly-granular siting problem is solved in order to identify a suitable subset of candidate sites where offshore wind power plants could be deployed. Then, in the second stage, the subset of locations selected in the first stage is passed to a capacity expansion planning framework that sizes the power generation, transmission and storage assets that should be deployed and operated in order to satisfy pre-specified electricity demand levels at minimum cost subject to technical and policy constraints. An open source tool implementing the two-stage method is also provided for the sake of transparency.}

\review{Two types of deployment schemes that select sites so as to maximise their aggregate power output and spatiotemporal complementarity are analysed. Roughly speaking, sites are considered complementary if they rarely experience simultaneous low electricity production events \cite{Berger2018}. Two variants of these siting schemes are also studied, wherein the number of sites to be selected is specified on a country-by-country basis rather than Europe-wide. A few hundred sites are identified using each scheme, a high resolution grid and ten years of reanalysis data \cite{ERA5}. These sites are then passed to a capacity expansion planning framework relying on a stylised model of the European power system where each country corresponds to an electrical bus and including an array of power generation and storage technologies. The framework sizes gas-fired power plants, offshore wind power plants, battery storage and electricity transmission assets and operates the system so as to supply electricity demand levels consistent with current European electricity consumption at minimum cost while reducing carbon dioxide emissions from the power sector by 90\% compared with 1990 levels and taking a broad range of legacy assets into account. A detailed sensitivity analysis is also performed in order to evaluate the impact of offshore wind cost assumptions and inter-annual weather variability on system design and economics.}

This manuscript is organised as follows. Section \ref{RelatedWorks} reviews the relevant studies in the literature. Section \ref{Model} presents the two-stage method used to evaluate the impact of RES siting strategies on power system design and economics. The case study is introduced in Section \ref{TestCase}, while results are presented and analysed in Section \ref{Results}. Section \ref{Conclusion} concludes the paper and discusses future work avenues.

\section{Related Works}\label{RelatedWorks}

\review{The precise estimation of required capacities and incurred costs in RES-dominated power systems relies heavily on a detailed modelling of RES assets \cite{Pfenninger2014}. In one of the first studies to quantify the impact of this modelling aspect, Krishnan and Cole \cite{Krishnan2016} reveal that using 356 and 134 profiles to model the wind and solar resource, respectively, within the contiguous US leads to significantly different capacity outcomes compared to the case where the same resources are modelled via one single profile per state (i.e., 48 profiles in total). For example, the solar PV capacity difference between the two set-ups exceeded \SI{32}{GW}, or 10\% of the total installed capacity for this technology. In a more recent assessment, Frysztacki et al. \cite{Brown2021} evaluate the role of high-resolution RES siting in a study focusing on the European power system. They confirm the findings of \cite{Krishnan2016} regarding the considerable impact of RES representation on the installed capacity requirements and, in addition, point out that modelling RES via 1024 different profiles leads to 10.5\% lower system costs compared to more simplified set-ups using only 37 distinct profiles (i.e., one per country). In the following, the detailed representation of RES to produce an accurate (i.e., high-resolution) assessment of the most suitable locations for asset deployment will be referred to as \textit{siting}. A set-up where the detailed RES representation is integrated in models whose outcomes include installed capacities and associated costs, will be referred to as \textit{sizing}.}

The siting of RES assets has been a growing research topic lately. The models tackling this problem typically put more emphasis on the representation of renewable resources at the expense of other modelling features such as network or time-coupling constraints. In addition, they use non-monetary objectives such as, e.g., residual load or resource variability minimization. For instance, Jerez et al. \cite{Jerez2015} propose a tool which enables the distribution of RES capacities, as well as their output estimation via several transfer functions, across a regular grid with a spatial resolution of 0.44\degree. The problem is tackled by first computing distribution keys that take into account resource quality, population density and the existence of protected areas and then leveraging them to spread pre-defined capacities of RES across the system. Becker and Thrän \cite{Becker2018} propose a method that sites wind generators such that the correlation (estimated via the Pearson coefficient) of the underlying resource with that of existing assets is minimized. A heuristic is also designed to solve the problem. Then, \review{Musselman et al. \cite{Musselman2019} tackle the wind farm siting problem via two different bi-objective models formulated as mixed-integer linear programs (MILP). The first model seeks to simultaneously minimize i) the average residual demand and ii) the average power output variability (measured as the absolute change in residual demand between consecutive periods), while the second model simultaneously minimizes i) the average residual demand and ii) the maximum increase in residual demand over a set of time periods of pre-specified length. Furthermore, a three-stage quadratic program (QP) proposed by Hu et al. \cite{Hu2019} uses portfolio optimisation concepts to site RES assets such that the standard deviation of their aggregate feed-in (i.e., the portfolio volatility) is minimized.} However, overlooking the electricity demand in this process brings into question the ability of some of these methods \cite{Becker2018, Hu2019} to achieve proper siting. A framework siting RES assets such that the occurrence of simultaneous, system-wide low-generation events is minimized has been recently proposed by Berger et al. \cite{Berger2018}. The problem has since been cast as an integer program (IP) for which several solution methods have been proposed \cite{Berger2020}. Although they offer a valuable overview of different siting criteria proposed in the literature, a common drawback of all these studies is that they fall short in evaluating the implications of the corresponding outcomes on the design and economics of power systems. Such a feature usually surfaces once a sizing model reveals the configuration of the power system.

The sizing of renewable power generation plants has been traditionally achieved via capacity expansion planning (CEP) frameworks, a class of problems which has received a great deal of attention in recent years \cite{Dagoumas2018}. \review{For example, Baringo and Conejo \cite{Baringo2014} have studied the strategic investment in wind power generation assets by making use of a bi-level formulation in which investment decisions (siting and sizing) define the upper level and market clearing forms the lower level problem. In addition, Munoz et al. \cite{Munoz2014} tackled the joint generation and transmission expansion planning via a MILP where investment decisions are done in two stages, such that corrective actions are possible once uncertainty is revealed. In theory, such models are capable of evaluating the implications of RES siting on the design and economics of power systems. However, owing to computational limitations, these models usually have relatively low spatial and temporal resolutions, an aspect that makes it difficult to accurately capture correlations between variable renewable resources and properly site RES assets. Several attempts to integrate spatially-resolved siting of RES assets have been made, yet a common drawback can be identified across all of them.} On the one hand, in line with \cite{Baringo2014, Munoz2014} where purely economic criteria are optimised, a study by MacDonald et al. \cite{MacDonald2016} leverages a CEP framework cast as a linear program (LP) to jointly optimise generation, transmission and storage capacities. The model is instantiated with hourly-sampled RES and demand data, while a \SI{13}{km} regular grid is used for an accurate representation of renewable resources. However, the approach is reported to require thousands of core hours to solve large-scale instances, a feature which makes it difficult to reproduce and limits its use in practice. Another study making use of a cost-minimization CEP framework cast as an LP \cite{Zappa2019} sites RES assets over a 0.75\degree \ regular grid. This time, the formulation of the CEP problem relies on a highly simplified temporal representation of the renewable resource availability (i.e., the hourly resolution is replaced by a 144-step duration curve), an aspect that often limits the ability of the underlying model to accurately estimate system needs \cite{Kotzur2018}. On the other hand, non-economic optimisation criteria have also been used in sizing set-ups. For instance, Wu et al. \cite{Callaway2017} propose an IP for siting and sizing wind generation at high spatial resolutions (e.g., \SI{3.6}{km} used in their study) such that the need for peak conventional generation feed-in is minimised. Nevertheless, their formulation, which relies on a full coefficient matrix, is computationally inefficient and its scalability is limited to a few hundred locations and one year of data with hourly resolution. In a similar fashion, Zappa and van den Broek \cite{Zappa2018} minimize year-round residual demand through a linearly constrained QP. In the proposed model, RES assets are sited over the same regular grid used in \cite{Zappa2019}. However, their method suffers from similar scalability issues as \cite{Callaway2017}, which limits the scale of problems tackled to a few hundred of RES sites and one year with hourly resolution.

\review{In this paper, thousands of sites are available as candidate deployment locations in the problem of siting offshore wind across European Seas. A two-stage routine that bridges the gap between the streams of literature independently tackling the siting and sizing of RES assets in CEP frameworks is then leveraged to assess the impact of different siting strategies on the design and economics of power systems.}

\section{Methodology}\label{Model}

This section describes the two-stage method combining asset siting schemes and a capacity expansion planning model. Some basic notation used throughout this section is first introduced. The models and solution methods used in the siting stage are discussed next. Finally, the capacity expansion planning framework is presented.

\subsection{Preliminaries}\label{Preliminaries}

A finite time horizon $T \in \mathbb{N}$ and associated set of time periods $\mathcal{T} = \{1, \ldots, T\}$ are considered. A geographical area is represented by a finite set of locations $\mathcal{L}, |\mathcal{L}| = L$, which may be partitioned into a collection of disjoint regions $\mathcal{L}_n \subseteq \mathcal{L}, \mbox{ } \forall n \in \mathcal{N}_B$, where $\mathcal{N}_B, |\mathcal{N}_B| = B,$ may for instance represent the set of electrical buses in a power system and $\mathcal{L}_n$ may represent a set of candidate RES sites that may be connected to bus $n \in \mathcal{N}_B$. \review{Each location $l \in \mathcal{L}$ is assumed to have a fixed technical potential $\Bar{\kappa}_l \in \mathbb{R}_+$, which represents the maximum capacity that may be deployed at this location. In addition, some legacy capacity $\underline{\kappa}_l \in \mathbb{R}_+$ may have already been deployed at sites $l \in \mathcal{L}_0 \subseteq \mathcal{L}$.} A time series $\mathbf{s}_l = \begin{pmatrix} s_{l1}, \ldots, s_{lT} \end{pmatrix} \in \mathbb{R}_+^T$ describing renewable resource data (e.g., wind speed, solar irradiation) over $\mathcal{T}$ is assumed to be available at each location $l \in \mathcal{L}$. Furthermore, the instantaneous power output of location $l \in \mathcal{L}$ is estimated via a suitable transfer function $h_l$ that returns per-unit capacity factor values $\pi_{lt} = h_l(s_{lt}), \forall t \in \mathcal{T}$, which are stored in a time series $\boldsymbol{\pi}_l = \begin{pmatrix} \pi_{l1}, \ldots, \pi_{lT} \end{pmatrix} \in \{0, 1\}^T$. This transfer function may be that of a single RES power generation technology (e.g., a wind turbine or a solar PV module) or that of an entire power station (e.g., a wind farm or a PV power station).

\subsection{Siting Schemes}\label{SitingModel}
The models and solution methods used in asset siting schemes are described in this section.

\subsubsection{Models}\label{ModelsPRODCOMP}

Models that select a pre-specified number of suitable candidate sites so as to optimise a given criterion are introduced. Two different criteria are considered, leading to two different siting schemes. The first criterion measures the aggregate power output ($PROD$), while the second one measures the spatiotemporal complementarity that sites exhibit ($COMP$). Both siting problems are cast as integer programming models.

\paragraph{Aggregate Power Output} \review{This siting scheme selects a collection of disjoint subsets of locations so as to maximise their average capacity factor. More precisely, a pre-specified number of locations $k_n \in \mathbb{N}$ (including legacy locations) is selected in each region $\mathcal{L}_n$, and the total number of locations that must be deployed is $k = \sum_{n \in \mathcal{N}_B} k_n$. In order to formulate this problem as an integer program, a set of binary variables is introduced. Indeed, a binary variable $x_l \in \{0, 1\}$ is used to indicate whether location $l$ is selected for deployment, that is, $x_l = 1$ if location $l$ is selected for deployment and $x_l = 0$ otherwise. A binary matrix with entries $A_{nl} \in \{0, 1\}$ is also used to indicate whether location $l$ belongs to region $\mathcal{L}_n$, such that $A_{nl} = 1$ if this is the case and $A_{nl} = 0$ otherwise. Note that since regions are disjoint, each column of this matrix has exactly one nonzero entry, and we may assume without loss of generality that locations are ordered such that matrix $A$ is block diagonal. The problem at hand then reads}

\begin{subequations}
\label{PROD}
\begin{align}
\label{ObjectivePROD}
\underset{x_l} \max \hspace{10pt} & \frac{1}{k}\sum_{l \in \mathcal{L}} x_l \bigg[\frac{1}{T} \sum_{t \in \mathcal{T}} \pi_{lt} \bigg] \\ \label{PartitioningPROD}
\mbox{s.t. } & \sum_{l \in \mathcal{L}} A_{nl} x_l = k_n, \mbox{ } \forall n \in \mathcal{N}_B,\\ \label{LegacyPROD}
& x_l = 1, \mbox{ } \forall l \in \mathcal{L}_0,\\ \label{DeploymentIntegralityPROD}
& x_l \in \{0, 1\},\mbox{ } \forall l \in \mathcal{L}.
\end{align}
\end{subequations}

\review{The objective function (\ref{ObjectivePROD}) computes the average capacity factor of the locations selected for deployment. The cardinality constraints (\ref{PartitioningPROD}) ensure that exactly $k_n$ locations are selected in region $\mathcal{L}_n$, $\forall n \in \mathcal{N}_B$, while constraints (\ref{LegacyPROD}) guarantee that legacy assets are taken into account. Finally, constraints (\ref{DeploymentIntegralityPROD}) express the binary nature of location selection decisions.}

\paragraph{Spatiotemporal Complementarity} This siting scheme selects a collection of disjoint subsets of locations so as to maximise their spatiotemporal complementarity. Recall that, in this paper, locations are considered complementary if they rarely experience simultaneous low electricity production events (compared with a pre-specified reference production level) \cite{Berger2018, Berger2020}. The framework that makes it possible to cast this problem as an integer program is discussed next.

First, a set of time windows $\mathcal{W}$, $|\mathcal{W}| = W$, is constructed from the set of time periods $\mathcal{T}$. More precisely, a time window $\mathrm{w} \in \mathcal{W}$ can be seen as a subset $\mathcal{T}_{\mathrm{w}} \subseteq \mathcal{T}$ of $\delta$ successive time periods, and all time windows $\mathrm{w} \in \mathcal{W}$ have the same length $\delta$. Note that successive time windows overlap and share exactly $\delta - 1$ time periods, while the union of all time windows covers the set of time periods $\mathcal{T}$. Then, the per-unit power generation level $\bar{\pi}_{l\mathrm{w}} \in [0, 1]$ of each candidate site $l \in \mathcal{L}$ is evaluated over the duration of each time window $\mathrm{w} \in \mathcal{W}$ using a prescribed measure $g_l$, such that $\bar{\pi}_{l\mathrm{w}} = g_l(\{\pi_{lt}\}_{t \in \mathcal{T}_\mathrm{w}})$. This measure may for instance compute the average production level over each window $\mathrm{w} \in \mathcal{W}$. This would essentially be equivalent to applying a moving average-based filter to the original power production signal and result in a smoothed power output signal. The degree of smoothing would be controlled by $\delta$, which makes it possible to study resource complementarity on different time scales. A local, time-dependent reference production level $\alpha_{l\mathrm{w}} \in \mathbb{R}_+$ is also specified at each candidate site $l \in \mathcal{L}$, and may for instance be proportional to the electricity demand. A location $l \in \mathcal{L}$ is considered productive enough over window $\mathrm{w}$ if $\bar{\pi}_{l\mathrm{w}} \ge \alpha_{l\mathrm{w}}$. Location $l$ is then said to \textit{cover} window $\mathrm{w}$ and be \textit{non-critical}. Checking whether this condition is satisfied for all locations and time windows enables the construction of a binary matrix with entries $D_{l\mathrm{w}} \in \{0, 1\}$, such that $D_{l\mathrm{w}} = 1$ if location $l$ covers window $\mathrm{w}$ and $D_{l\mathrm{w}} = 0$ otherwise. In order to formalise the intuitive definition of resource complementarity introduced earlier, a threshold $c \in \mathbb{N}$ is specified, such that for any subset of candidate locations $L \subseteq \mathcal{L}$, a window $\mathrm{w} \in \mathcal{W}$ is said to be \textit{$c$-covered} or \textit{non-critical} if at least $c$ locations cover it (i.e., produce enough electricity over its duration). More formally, window $\mathrm{w}$ is non-critical if $\sum_{l \in L} D_{l\mathrm{w}} \ge c$. 

Using the notation introduced for the first siting scheme, formulating the integer programming problem only requires the definition of a set of additional binary variables. More precisely, for each window $\mathrm{w} \in \mathcal{W}$, a binary variable $y_\mathrm{w} \in \{0, 1\}$ indicating whether window $\mathrm{w}$ is non-critical is introduced, such that $y_\mathrm{w} = 1$ if window $\mathrm{w}$ is non-critical and $y_\mathrm{w} = 0$ otherwise. The problem of siting renewable power plants so as to maximise their spatiotemporal complementarity then reads

\begin{subequations}
\label{COMP}
\begin{align}
\label{ObjectiveCOMP}
\underset{x_l, y_\mathrm{w}} \max \hspace{10pt} & \sum_{\mathrm{w} \in \mathcal{W}} y_\mathrm{w}\\ \label{CriticalityDefinitionConstraintCOMP}
\mbox{s.t. } & \sum_{l \in \mathcal{L}} D_{l\mathrm{w}} x_l \ge c y_\mathrm{w}, \mbox{ } \forall \mathrm{w} \in \mathcal{W}, \\ \label{PartitioningCOMP}
& \sum_{l \in \mathcal{L}} A_{nl} x_l = k_n, \mbox{ } \forall n \in \mathcal{N}_B,\\ \label{LegacyCOMP}
& x_l = 1, \mbox{ } \forall l \in \mathcal{L}_0,\\ \label{DeploymentIntegralityCOMP}
& x_l \in \{0, 1\},\mbox{ } \forall l \in \mathcal{L}, \\  \label{WindowIntegralityCOMP}
& y_\mathrm{w} \in \{0, 1\}, \mbox{ } \forall \mathrm{w} \in \mathcal{W}.
\end{align}
\end{subequations}

The objective function (\ref{ObjectiveCOMP}) simply computes the number of non-critical time windows observed over the time horizon of interest. Dividing the objective by the total number of time windows shows that it can be interpreted as quantifying the empirical probability of having sufficient levels of electricity production across at least $c$ locations simultaneously. A low objective value therefore implies that simultaneous low electricity production events occur often, which indicates poor complementarity between locations. Constraints (\ref{CriticalityDefinitionConstraintCOMP}) define the binary classification of time windows and express the fact that a time window $\mathrm{w} \in \mathcal{W}$ is non-critical if at least $c$ locations selected for deployment cover it. The cardinality constraints (\ref{PartitioningCOMP}) ensure that exactly $k_n$ locations are selected in region $\mathcal{L}_n$, $\forall n \in \mathcal{N}_B$, while constraints (\ref{LegacyPROD}) guarantee that legacy assets are accounted for in siting decisions. Finally, constraints (\ref{DeploymentIntegralityCOMP}-\ref{WindowIntegralityCOMP}) express the binary nature of location selection decisions and time window criticality, respectively.

\subsubsection{Solution Methods}\label{SolutionMethod}

The solution methods used to tackle problems (\ref{PROD}) and (\ref{COMP}) are discussed next.

\paragraph{Aggregate Power Output} \review{Since the objective function (\ref{ObjectivePROD}) is separable and the coefficient matrix in Eq. (\ref{PartitioningPROD}) is block diagonal, problem (\ref{ObjectivePROD}-\ref{DeploymentIntegralityPROD}) is straightforward to decompose and solve. More precisely, the $k_n$ most productive locations can be selected independently in each region $\mathcal{L}_n$, $\forall n \in \mathcal{N}_B$. In each region $\mathcal{L}_n$, this can be achieved by successively i) computing the average capacity factor of each location $\tilde{\pi}_l = (1/T) \sum_{t \in \mathcal{T}} \pi_{lt}, \mbox{ }\forall l \in \mathcal{L}_n$, ii) sorting locations based on their average capacity factor $\tilde{\pi}_l$, iii) adding the locations with the highest average capacity factors to a set $L_n \subseteq \mathcal{L}_n$ that initially contains the legacy locations belonging to region $n$, until $|L_n| = k_n$. The solution $L$ to problem (\ref{PROD}) is then obtained by taking the union of these sets, $L = \bigcup_{n \in \mathcal{N}_B} L_n$.}

\paragraph{Spatiotemporal Complementarity} An approximate solution method relying on a mixed-integer relaxation of problem (\ref{COMP}) followed by a local search algorithm inspired by the simulated annealing algorithm \cite{Bertsimas1993} is used to tackle (\ref{ObjectiveCOMP}-\ref{WindowIntegralityCOMP}) \cite{Berger2020}. 

The mixed-integer relaxation is formed by relaxing the integrality constraint \eqref{WindowIntegralityCOMP} of the time window variables. The key advantage of this approach lies in the fact that the siting variables $x_l, \forall l \in \mathcal{L}$, remain integer in the solution and subsets of locations $L_n \subseteq \mathcal{L}_n, \mbox{ } \forall n \in \mathcal{N}_B$, can be directly extracted from them. The number of non-critical windows associated with this collection of subsets $L = \bigcup_{n \in \mathcal{N}_B} L_n$ can be computed via a function $f_c$ such that
\begin{equation}
f_c(L) = \Big|\Big\{\mathrm{w} \in \mathcal{W} \Big | \sum_{l \in L} D_{\mathrm{w}l} \ge c \Big\}\Big|.
\label{setfunction}
\end{equation}

The local search algorithm starts from a subset of locations $L_0 \subseteq \mathcal{L}$ obtained by solving the mixed-integer relaxation of problem (\ref{COMP}). \review{Note that by construction, $L_0$ includes legacy locations and satisfies the cardinality constraints (\ref{PartitioningCOMP}). Since legacy locations cannot change, they are first removed from $L_0$ in order to initialise the incumbent solution $L \subseteq \mathcal{L}$. Likewise, legacy locations are removed from the set of candidate sites that may be selected in each region $\mathcal{L}_n, \mbox{ } \forall n \in \mathcal{N}_B$. Then, the algorithm performs a fixed number of iterations $I \in \mathbb{N}$ in the hope of improving the incumbent solution. More specifically, in each iteration, a fixed number $N \in \mathbb{N}$ of neighbouring solutions is drawn at random from the neighbourhood of the incumbent solution. This neighbourhood is formed by solutions that satisfy the cardinality constraints (\ref{PartitioningCOMP}) and share exactly $k - r$ locations with the incumbent solution. A neighbouring solution $\hat{L}$ can be constructed from the incumbent solution as follows. For each region $\mathcal{L}_n$, $\mathbf{s}(n)$ different locations are sampled uniformly at random from both $\mathcal{L}_n \setminus L$ and $\mathcal{L}_n \cap L$, and these locations are swapped. The numbers of locations sampled in different regions are chosen at random such that the cardinality constraints (\ref{PartitioningCOMP}) remain satisfied and $\sum_{n \in \mathcal{N}_B} \mathbf{s}(n) = r$. Then, each of the $N$ neighbouring solutions is tested against the incumbent solution and stored in a temporary candidate solution $\tilde{L}$ if it is found to outperform previously-explored neighbouring solutions. Their performance is evaluated via the difference $\tilde{\Delta}$ between the objectives achieved by the neighbouring and incumbent solutions. Once $N$ neighbouring solutions have been explored, the candidate solution corresponds to a neighbouring solution that maximises $\tilde{\Delta}$ among all sampled solutions. Note that $\tilde{\Delta}$ may be negative (i.e., if the algorithm does not manage to improve on the incumbent). If $\tilde{\Delta} > 0$, the candidate solution becomes the new incumbent solution. By contrast, if $\tilde{\Delta} < 0$, whether the candidate solution becomes the new incumbent solution depends on the outcome $b$ of a random variable drawn from a Bernoulli distribution with parameter $p$. This parameter depends on both $\tilde{\Delta}$ and the so-called \textit{annealing temperature} $T(i)$. Roughly speaking, the temperature controls the extent to which the search space is explored in an attempt to find better solutions and exit local optima. The temperature is specified by a \textit{temperature schedule} that provides a temperature $T(i)$ for each iteration $i$. This procedure is repeated until the maximum number of iterations $I$ is reached. Algorithm \ref{SA} summarises these ideas.}

\begin{algorithm}
	\caption{Local Search Algorithm}\label{SA}
	\begin{algorithmic}[1]
		\Require $L_0, \mbox{ } \mathcal{L}_0, \mbox{ } \mathcal{N}_B, \mbox{ } \{\mathcal{L}_n\}_{n \in \mathcal{N}_B}, \mbox{ } I, \mbox{ } N,  \mbox{ } r, \mbox{ } T, \mbox{ } f_c$
		\State $L \gets L_0 \setminus \mathcal{L}_0$
		\For{$n \in \mathcal{N}_B$}
		\State $\mathcal{L}_n \gets \mathcal{L}_n \setminus \mathcal{L}_0$
		\EndFor
		\State $i \gets 0$
		\While{$i < I$}
		\State $\tilde{\Delta} \gets -\infty$
		\State $j \gets 0$
		\While{$j < N$}
		\State $\hat{L} \gets L$
		\State $\mathbf{s} \gets \mbox{vector storing the number of locations to sample per region} $
		\For{$n \in \mathcal{N}_B$}
		\State $S_+ \gets \mathbf{s}(n) \mbox{ locations sampled from } \mathcal{L}_n \setminus L \mbox{ uniformly at random}$
		\State $S_- \gets \mathbf{s}(n) \mbox{ locations sampled from } \mathcal{L}_n \cap L \mbox{ uniformly at random}$
		\State $\hat{L} \gets (\hat{L} \setminus S_-) \cup S_+$
		\EndFor
		\State $\hat{\Delta} \gets f_c(\hat{L} \cup \mathcal{L}_0) - f_c(L \cup \mathcal{L}_0)$
		\If{$\hat{\Delta} > \tilde{\Delta}$}
		\State $\tilde{L} \gets \hat{L}$
		\State $\tilde{\Delta} \gets \hat{\Delta}$
		\EndIf
		\State $j \gets j+1$
		\EndWhile
		\If{$\tilde{\Delta} > 0$}
		\State $L \gets \tilde{L}$
		\Else
		\State $p \gets \mbox{$\mathrm{exp}$}(\tilde{\Delta} / T(i))$
		\State $\mbox{draw } b \mbox{ from Bernoulli distribution with parameter $p$}$
		\If{$b = 1$}
		\State $L \gets \tilde{L}$
		\EndIf
		\EndIf
		\State $i \gets i+1$
		\EndWhile
		\State $L \gets L \cup \mathcal{L}_0$
		\Ensure $L, f_c(L)$
	\end{algorithmic}
\end{algorithm}

\subsection{Capacity Expansion Planning Framework}\label{SizingModel}

Upon retrieving a suitable subset of locations $L \subseteq \mathcal{L}$ from the siting stage, the associated capacity factor time series $\{\boldsymbol{\pi}_l\}_{l \in L}$, legacy capacities $\{\underline{\kappa}_l\}_{l \in L}$ and technical potentials $\{\bar{\kappa}_l\}_{l \in L}$ are passed as input data to a capacity expansion planning (CEP) framework that determines the optimal power system design. More precisely, the CEP model described in (\ref{eq:objective_sizing}-\ref{eq:prm_constraint}) selects and sizes the power generation, transmission and storage assets that should be deployed and operated in order to satisfy pre-specified electricity demand levels across Europe at minimum cost subject to a set of technical and environmental constraints. \review{In the formulation below, Latin letters denote optimization variables, while Greek characters represent problem parameters.}

A set of working assumptions characterize the capacity expansion planning framework used in the current study. First, investment decisions in power system assets are made by a central planner who also operates the system, has perfect foresight, and whose goal is to minimise total system cost in a purely deterministic set-up. A static investment horizon is considered and the investment and operation problems are solved concurrently. Investment decisions are made once (at the beginning at the optimisation horizon), while the operational decisions are taken on an hourly basis. Second, investments in generation, transmission or storage capacities are continuous variables and transmission expansion is limited to the reinforcement of existing corridors. Third, the network is represented by i) a set of existing nodes, which represent an aggregation of real electrical nodes and ii) a set of existing transmission corridors, which connect the aforementioned nodes. Legacy generation assets at existing nodes are taken into account. Additional dispatchable capacity (e.g., gas-fired power plants) may be added at those nodes, while additional RES generation capacity may also be built at existing nodes, provided that the local renewable potential is not fully exploited. Finally, as seen in (\ref{eq:objective_sizing}-\ref{eq:prm_constraint}), no unit commitment constraints are considered and the full operating range of dispatchable power plants is assumed stable.

The objective function (\ref{eq:objective_sizing}) includes the capacity-dependent investment and fixed operation and maintenance (O\&M) costs and the output-dependent variable O\&M expenditures. In addition, an economic penalty is enforced on electricity demand shedding. Then, the electricity supply and demand balance is enforced via (\ref{eq:energy_balance}).

\allowdisplaybreaks
\begin{subequations}
\begin{align}
&\underset{\tiny \begin{array}{cc} \mathbf{K}, (\mathbf{p}_t)_{t \in \mathcal{T}} \end{array}} \min \hspace{2mm} \sum_{\substack{n \in \mathcal{N}_B \\ l \in L_n}} \big(\zeta^{l} + \theta^l_f\big) K_{l} + \sum_{\substack{n \in \mathcal{N}_B \\ j \in \mathcal{G}\cup\mathcal{R}\cup\mathcal{S}}} \big(\zeta^{j} + \theta^j_f\big) K_{nj} + \sum_{\substack{n \in \mathcal{N}_B \\ s \in \mathcal{S}}} \zeta^{s}_S S_{ns} + \sum_{c \in \mathcal{C}} \big(\zeta^c + \theta^c_f\big) K_c \label{eq:objective_sizing}\\\nonumber & + \sum_{t \in \mathcal{T}} \omega_t \Big[\sum_{\substack{n \in \mathcal{N}_B \\ l \in L_n}} \theta^l_v p_{lt} + \sum_{\substack{n \in \mathcal{N}_B \\ g \in \mathcal{G}\cup\mathcal{R}}} \theta^g_v p_{ngt} + \sum_{\substack{n \in \mathcal{N}_B \\ s \in \mathcal{S}}} \theta^s_v (p^C_{nst} + p^D_{nst}) + \sum_{c \in \mathcal{C}} \theta^c_v \abs{p_{ct}} + \sum_{n \in \mathcal{N}_B} \theta^{ens} p^{ens}_{nt} \Big]\\[2pt] 
\text{s.t.} &\sum_{\substack{n \in \mathcal{N}_B \\ l \in L_n}} p_{lt} + \sum_{g \in \mathcal{G}\cup\mathcal{R}} p_{ngt} + \sum_{s \in \mathcal{S}} p^D_{nst} + \sum_{c \in \mathcal{C}_n^{+}} p_{ct} + p^{ens}_{nt} \label{eq:energy_balance}\\\nonumber & \hspace{30mm} = \lambda_{nt} + \sum_{s \in \mathcal{S}} p^C_{nst} + \sum_{c \in \mathcal{C}_n^{-}} p_{ct} , \mbox{ } \forall n \in \mathcal{N}_B, \forall t \in \mathcal{T}\\[2pt] \intertext{The operation and deployment of the RES units whose locations are determined by leveraging the siting models in Section \ref{SitingModel} are constrained by (\ref{eq:res_flow_definition}) and (\ref{eq:res_sizing_limit}), respectively. The next six equations model the operation and sizing of the remaining generation units, including RES technologies that are not sited. More specifically, modelling aspects such as instantaneous in-feed (\ref{eq:dispatchable_flow_definition}), ramp rates (\ref{eq:dispatchable_positive_rate}-\ref{eq:dispatchable_negative_rate}), minimum operating levels (\ref{eq:dispatchable_min_level}), $\mathrm{CO_2}$ emission levels (\ref{eq:dispatchable_co2_definition}) or technical potential limitations (\ref{eq:dispachable_sizing_limit}) are considered.}
&p_{lt} \le \pi_{lt} (\underline{\kappa}_{l} + K_{l}), \mbox{ } \forall l \in L_n, \forall n \in \mathcal{N}_B, \forall t \in \mathcal{T} \label{eq:res_flow_definition}\\[2pt]
&\underline{\kappa}_{l} + K_{l} \le \Bar{\kappa}_{l}, \mbox{ } \forall l \in L_n, \forall n \in \mathcal{N}_B \label{eq:res_sizing_limit} \\[2pt] 
&p_{ngt} \le \pi_{nt} (\underline{\kappa}_{ng} + K_{ng}), \mbox{ } \forall n \in \mathcal{N}_B, \forall g \in \mathcal{G}, \forall t \in \mathcal{T} \label{eq:dispatchable_flow_definition} \\[2pt]
&p_{ngt} - p_{ngt-1} \le \Delta_g^{+} (\underline{\kappa}_{ng} + K_{ng}), \label{eq:dispatchable_positive_rate} \forall n \in \mathcal{N}_B, \forall g \in \mathcal{G}, \forall t \in \mathcal{T} \setminus \{0\}\\[2pt]
&p_{ngt} - p_{ngt-1} \ge -\Delta_g^{-} (\underline{\kappa}_{ng} + K_{ng}), \label{eq:dispatchable_negative_rate} \forall n \in \mathcal{N}_B, \forall g \in \mathcal{G}, \forall t \in \mathcal{T} \setminus \{0\}\\[2pt]
&\mu_g (\underline{\kappa}_{ng} + K_{ng}) \le p_{ngt}, \mbox{ } \forall n \in \mathcal{N}_B, \forall g \in \mathcal{G}, \forall t \in \mathcal{T} \label{eq:dispatchable_min_level} \\[2pt]
&q^{CO_2}_{ngt} = \nu^{CO_2}_{g} p_{ngt} / \eta_{g}, \mbox{ } \forall n \in \mathcal{N}_B, \forall g \in \mathcal{G}, \forall t \in \mathcal{T} \label{eq:dispatchable_co2_definition} \\[2pt] 
&\underline{\kappa}_{ng} + K_{ng} \le \Bar{\kappa}_{ng}, \mbox{ } \forall n \in \mathcal{N}_B, \forall g \in \mathcal{G} \label{eq:dispachable_sizing_limit} \\[2pt] \intertext{\review{It is worth mentioning the two most common situations in which the latter six constraints are enforced. On the one hand, if dispatchable units are modelled (e.g., gas-fired power plants), the time-dependent availability $\pi_{nt}$ in Eq. (\ref{eq:dispatchable_flow_definition}) is set to one across the entire optimisation horizon. On the other hand, if a RES technology not sited via the models in Section \ref{SitingModel} is addressed, the aforementioned parameter is instantiated with a per-unit capacity factor time series that is aggregated at the spatial resolution represented by bus $n \in \mathcal{N}_B$. Furthermore, the per-unit ramp rates $\Delta_g^{+}$ and $\Delta_g^{-}$ in Eq. (\ref{eq:dispatchable_positive_rate}-\ref{eq:dispatchable_negative_rate}) are set to one, while the must-run and the specific $\mathrm{CO_2}$ emission levels $\mu_g$ and $\nu^{CO_2}_{g}$, respectively, are set to zero.}}
&p^D_{nst} \le K_{ns}, \mbox{ } \forall n \in \mathcal{N}_B, \forall s \in \mathcal{S}, \forall t \in \mathcal{T} \label{eq:storage_discharge_definition} \\[2pt]
&p^C_{nst} \le \phi_s K_{ns}, \mbox{ } \forall n \in \mathcal{N}_B, \forall s \in \mathcal{S}, \forall t \in \mathcal{T} \label{eq:storage_charge_definition} \\[2pt]
&e_{nst} = \eta^{SD}_s e_{nst-1} + \omega_s \eta^{C}_s p^C_{nst} - \omega_s \frac{1}{\eta^D_{s}} p^D_{nst}, \forall n \in \mathcal{N}_B, \forall s \in \mathcal{S}, \forall t \in \mathcal{T}\label{eq:storage_soc_definition}\\[2pt]
&\mu_s S_{ns} \le e_{nst} \le S_{ns}, \forall n \in \mathcal{N}_B, \forall s \in \mathcal{S}, \forall t \in \mathcal{T}\label{eq:storage_soc_limits}\\[2pt]
&\underline{\kappa}_{ns} \le S_{ns} \le \Bar{\kappa}_{ns}, \mbox{ } \forall n \in \mathcal{N}_B, \forall s \in \mathcal{S} \label{eq:storage_capacity_limits} \\ \intertext{Storage units are modelled via (\ref{eq:storage_discharge_definition}) to (\ref{eq:storage_capacity_limits}), \review{assuming independent energy and power ratings and asymmetric charge/discharge rates}, while constraints (\ref{eq:transmission_flow_definition}) and (\ref{eq:transmission_capacity_limit}) define the transportation model governing the flow in transmission assets.}
&|p_{ct}| \le (\underline{\kappa}_c + K_c), \mbox{ } \forall c \in \mathcal{C}, \forall t \in \mathcal{T} \label{eq:transmission_flow_definition} \\[2pt]
&\underline{\kappa}_c + K_c \le \Bar{\kappa}_c, \mbox{ } \forall c \in \mathcal{C} \label{eq:transmission_capacity_limit} \\[2pt]
&\sum_{\substack{n \in \mathcal{N}_B \\ g \in \mathcal{G} \\ t \in \mathcal{T}}} q^{CO_2}_{ngt} \le \Psi^{CO_2} \label{eq:co2_budget_definition}\\
&\sum_{\substack{d \in \mathcal{D}}} K_{nd} + \sum_{\substack{r \in \mathcal{R}}} \Pi_{nr} K_{nr} + \sum_{\substack{l \in L_n}} \Pi_{l} K_{l} \ge (1+\Phi_n) \hat{\lambda}_{n}, \mbox{ } \forall n \in \mathcal{N}_B \label{eq:prm_constraint}
\end{align}
\end{subequations}
A system-wide $\mathrm{CO_2}$ budget is enforced via (\ref{eq:co2_budget_definition}). \review{Then, a system adequacy constraint is enforced via (\ref{eq:prm_constraint}) following the definition provided in \cite{Mertens2018}, according to which a system is adequate in the long-term by ensuring that the amount of firm capacity exceeds the peak demand by a planning reserve margin. According to Eq. (\ref{eq:prm_constraint}), this constraint is enforced at every bus $n \in \mathcal{N}_B$ and the corresponding peak demands and reserve margins are defined by $\hat{\lambda}_n$ and $\Phi_n$, respectively. There are two main sources providing firm capacity. On the one hand, set $\mathcal{D}$ in the first term on the left-hand side gathers dispatchable power generation technologies. On the other hand, RES assets also contribute to the provision of firm capacity and their participation is proportional to their capacity credit, as defined in \cite{Milligan2008}. To this end, two sets of RES technologies are defined. The one in the second term of (\ref{eq:prm_constraint}) gathers the subset of RES technologies which are not sited, while $L_n$ defines, for every $n \in \mathcal{N}_B$ the collection of sites obtained from the previous siting stage.}

\subsection{Implementation}
With the exception of the siting algorithm (detailed in Section \ref{SolutionMethod}) which was implemented in Julia 1.4, the implementation of the proposed framework is based on Python 3.7. All simulations were run on a workstation running under CentOS, with an 18-core Intel Xeon Gold 6140 CPU clocking at 2.3 GHz and 256 GB RAM. The sizing problem (\ref{eq:objective_sizing}-\ref{eq:prm_constraint}) is implemented in PyPSA 0.17 \cite{PyPSA}. Gurobi 9.1 was used to solve the MIR of (2), as well as (\ref{eq:objective_sizing}-\ref{eq:prm_constraint}). 

\section{Case Study}\label{TestCase}

The upcoming section describes the case study used to investigate i) the outcome of siting offshore wind plants within European borders by leveraging the two siting strategies introduced in Section \ref{SitingModel} and ii) the impact these siting strategies have on the resulting power system configurations. First, the realistic set-up used in the siting stage is presented. Then, the main features of the CEP framework are introduced. Recall that, in this exercise, offshore wind is the only renewable resource for which siting decisions are analysed, while the other RES technologies (i.e., onshore wind, utility-scale and distributed PV) are modelled via aggregate, per-country profiles obtained from the \textit{renewables.ninja} data platform \cite{Staffell2016Solar, Pffenninger2016Wind}.

\subsection{Offshore Wind Siting}\label{TestCaseSiting}

\paragraph{Renewable Resource Data} For this analysis, ten years (i.e., 2010 to 2019) of hourly-sampled wind speed data at a spatial resolution of \ang{0.25} are obtained from the ERA5 reanalysis dataset \cite{ERA5}. \review{The time series are then re-sampled by preserving the mean of each consecutive subset of three hours across the entire time horizon, yielding $T=29216$ time periods.} The conversion of raw resource data into capacity factor time series (a step required in both siting strategies introduced in Section \ref{SitingModel}) is achieved by applying the transfer function of a wind farm to the time series of wind speeds. \review{Determining the appropriate wind farm transfer function for each candidate site involves a two-step process. First, the ten-year average wind speed is computed and the relevant IEC wind class is determined \cite{IEC61400}. Once the wind class is known, an appropriate wind turbine is selected (in this exercise, two wind turbines are available, i.e., the Vestas V90 and V164 models) and the corresponding farm-specific transfer function is determined via a power curve smoothing procedure inspired from \cite{Holttinen2004}}.

\paragraph{Deployment Targets} Initially, an a priori filtering of candidate offshore wind locations is performed in order to discard sites where it would be impractical to deploy wind power plants. To this end, the following criteria are considered. First, a latitude threshold of 70\degree N is considered and all candidate locations beyond that limit are discarded. Second, sites with an average depth (i.e., the water depth across the entire reanalysis grid cell associated with the site) beyond \SI{999}{m} are also discarded. Third, only candidate sites situated between \SI{12}{nm} and \SI{120}{nm} (i.e., nautical miles) from the shore are further considered \cite{ENSPRESO2019}. \review{Finally, sites where operational or already planned offshore wind farms exist \cite{WindDataBase} are added back to the set of candidate locations, in case they were discarded following the application of the aforementioned filters.} As a result, a total of $L=2472$ candidate offshore sites (whose distribution per EEZ can be seen in the $|\mathcal{L}_n|$ column of Table \ref{tab:site_distributions}) are available in the siting stage. A visualization of the set of candidate sites is provided in the Supplementary Material.

In order to compute the number of sites $k$ that should be considered for deployment, the siting stage assumes the need for up to \SI{450}{GW} of offshore wind across 19 Exclusive Economic Zones (EEZ) within Europe, in line with a recent study published under the aegis of the European Commission \cite{WindEurope2020_main}. The conversion of the capacity requirements within each EEZ ($\kappa_n$) into the cardinality constraints ($k_n$) required in (\ref{ObjectivePROD}-\ref{DeploymentIntegralityPROD}) and (\ref{ObjectiveCOMP}-\ref{WindowIntegralityCOMP}) is achieved through Eq. (\ref{eq:capacity_to_cardinality}), where $\ceil*{\cdot}$ denotes the ceiling function, $\rho_{r}$ denotes the power density of the RES technology $r \in \mathcal{R}$ (expressed in $\mathrm{MW/km^2}$), $\sigma_{site}$ represents the surface area of a (generic) candidate location (expressed in $\mathrm{km^2}$) and $\epsilon_{site}$ denotes the dimensionless cell surface utilisation factor (since only a share of the cell surface area can be exploited for RES deployment purposes due to competing land uses). \review{Given that, at this stage, the geo-positioning of the $k$ sites to be identified is not known, average values for the last two parameters are considered. In particular, an offshore power density of $\SI{6}{MW/km^2}$ (consistent recent developments in the North Sea basin \cite{BLines2018}), a candidate site surface area of $\SI{442.5}{km^2}$ (corresponding to a \ang{0.25}-resolution cell at a latitude of 55\degree N) and a cell surface utilization factor of 50\% were considered.} In consequence, a total of 350 sites (whose per-country distribution can be seen in column $\tilde{k}_n$ of Table \ref{tab:site_distributions}) are required to accommodate the targeted \SI{450}{GW} of offshore wind.

\begin{align}\label{eq:capacity_to_cardinality}
\tilde{k}_n = \ceil*{\frac{\kappa_n}{\rho_{r} \times \sigma_{site} \times \epsilon_{site}}}, \mbox{ } \forall n \in \mathcal{N}_B
\end{align}

\review{According to a recent survey \cite{WindDataBase}, around \SI{99}{GW} of offshore wind capacity are currently in operation or in various stages of planning (e.g., construction, permitting, etc.) within European borders. When siting RES assets, taking into account the existence of legacy installations has a significant impact on the final outcome, as their location (and the corresponding generation patterns) influences the resulting deployment schemes. In the exercise at hand, the exact geo-positioning of the available legacy wind installations (retrieved from \cite{WindDataBase}) is mapped to the reanalysis grid used to perform the siting exercise, such that each wind farm is associated with an ERA5 grid point. This procedure reveals that, among the $L=2472$ candidate sites, $|\mathcal{L}_0| = 135$ of them (whose distribution per country is given in column $|\mathcal{L}_{0,n}|$ of Table \ref{tab:site_distributions}) have at least \SI{100}{MW} of legacy wind installations. At this point, a final adjustment is required to determine the values of $k_n$. This adjustment is necessary in two particular cases, i.e, i) when the number of legacy sites exceeds the previously computed $\tilde{k}_n$ or ii) when the number of candidate sites does not suffice to accommodate the required capacity under the aforementioned ($\rho_{r}, \sigma_{site}, \epsilon_{site}$) assumptions, and is enforced via Eq. (\ref{eq:capacity_to_cardinality_adjusted}).} As a result of this final adjustment, three additional sites (i.e., all of them associated with the EEZ of Greece) are added to the sets of required deployments and candidate locations, respectively. The resulting values for $k_n$ are gathered under the column with the same name in Table \ref{tab:site_distributions}. Throughout the analysis, both partitioned (i.e., $B=19$, where offshore wind sites are deployed whilst respecting the $k_n$ values per EEZ specified in Table \ref{tab:site_distributions}) and un-partitioned (i.e, $B=1$, where the $k = \sum_{n \in \mathcal{N}_B} k_n$ sites are freely deployed across the European Seas) siting strategies will be investigated.

\begin{align}\label{eq:capacity_to_cardinality_adjusted}
k_n = \min\{{|\mathcal{L}_n|, \max\{{|\mathcal{L}_{0,n}|,\tilde{k}_n\}}\}}, \mbox{ } \forall n \in \mathcal{N}_B
\end{align}

\begin{table}
\centering
\caption{Capacity requirements and cardinalities of various locations sets for the 19 European countries included in this study. Table entries sorted in descending order based on the capacity requirements per EEZ.}
\renewcommand{\arraystretch}{1.1}
\begin{tabular}{lccccc}
\toprule
EEZ & $\kappa_n$ & $|\mathcal{L}_n|$ & $|\mathcal{L}_{0,n}|$ & $\tilde{k}_n$ & $k_n$ \\
& [GW] & [sites] & [sites] & [sites] & [sites] \\
\midrule
UK & 80           & 700           & 39          & 61     &  61 \\
NL & 60           & 102           & 8          & 46      &  46 \\
FR & 57           & 231           & 7          & 43      &  43 \\
DE & 36           & 81            & 17          & 28     &  28 \\
DK & 35           & 119           & 15          & 27     &  27 \\
NO & 30           & 187           & 1          & 23      &  23 \\
PL & 28           & 51            & 10          & 22      &  22 \\
IE & 22           & 219           & 5          & 17      &  17 \\
IT & 20           & 112           & 2          & 16      &  16 \\
SE & 20           & 254           & 9          & 16       &  16 \\
FI & 15           & 128           & 5          & 12      &  12 \\
ES & 13           & 77            & 0          & 10      & 10 \\
GR & 10           & 39            & 11          & 8     &   11 \\
PT & 9            & 17            & 1           & 7       &  7 \\
BE & 6            & 4             & 2           & 5      &  4 \\
LV & 4            & 8             & 1           & 4      & 4 \\
LT & 3            & 49            & 0           & 3     &  3 \\
EE & 1            & 47            & 2           & 1     &  2 \\
HR & 1            & 47            & 0           & 1     &  1 \\ 
\bottomrule
\end{tabular}
\label{tab:site_distributions}
\end{table}

\paragraph{\textit{COMP} Siting Set-Up} \review{The $COMP$ siting strategy is carried out for a time window length $\delta$ of one time period (i.e., three hours). Then, a location $l \in \mathcal{L}$ is considered non-critical during time window $\mathrm{w}$ if its maximum theoretical generation potential exceeds a pre-defined share of the system-wide electricity demand. By expressing the former as the product between the technical potential $\Bar{\kappa}_l$ and the capacity factor $\Bar{\pi}_{l\mathrm{w}}$, this condition can be written as}
\begin{align}\label{eq:criticality_definition}
\Bar{\kappa}_l \Bar{\pi}_{l\mathrm{w}} \ge \frac{\varsigma \Bar{\lambda}_\mathrm{w}}{k} \ ,
\end{align}
\review{where $\varsigma$ represents the proportion of the electricity demand during window $\mathrm{w}$ (i.e, $\Bar{\lambda}_\mathrm{w}$) to be covered by offshore wind plants (which in this exercise is uniformly set to 30\%, as suggested in \cite{EC_2020}) and $k$ denotes the number of system-wide offshore wind deployments, whose per-partition distribution is detailed in Table \ref{tab:site_distributions}. Dividing both sides of Eq. (\ref{eq:criticality_definition}) by $\Bar{\kappa}_l$ yields the local criticality definition introduced in Section \ref{ModelsPRODCOMP}, where the reference production level $\alpha_{l\mathrm{w}} = \varsigma \Bar{\lambda}_{\mathrm{w}} / \Bar{\kappa}_l k$. Furthermore, threshold $c$ in Eq. (\ref{CriticalityDefinitionConstraintCOMP}) is set such that at least half of the locations must cover any time window for it to be labeled non-critical. In order to retrieve the $COMP$ set of sites, Algorithm \ref{SA} is run thirty times and the solution with the highest objective function (i.e., the highest number of non-critical windows) is retrieved and passed to the subsequent CEP stage. With respect to the algorithm parameters, a neighbourhood radius $r$ of 1, an initial temperature $T$ of 100 and an exponential temperature schedule $T(i) = 100 \times \exp(-10 \times i/I)$ were considered. Additionally, $I=5000$ iterations with $N=500$ neighbouring solutions each are considered for each run of the algorithm.}


\subsection{Capacity Expansion Problem}\label{TestCaseSizing}

\paragraph{Network Topology} The set of countries considered in the sizing stage includes, aside from the ones listed in Table \ref{tab:site_distributions}, Austria, Hungary, The Czech Republic, Slovakia, Switzerland (as landlocked territories), Bulgaria, Romania and Slovenia (with no offshore wind capacity mentioned in \cite{WindEurope2020_main}). It should be noted that $L_n = \emptyset$ for the subset of countries previously mentioned (i.e., no offshore wind sites available). Each country is modelled as one node, while the network topology is based upon that used for the 2018 version of the \textit{TYNDP} \cite{TYNDP2018}. A map of the topology is provided in the Supplementary Material. It is hereby assumed that all interconnections crossing bodies of water are developed as DC cables, while the remainder are AC cables. As mentioned previously, transmission expansion decisions are limited to the reinforcement of existing corridors. The connection costs of offshore sites to the associated onshore buses depend on the capacity of the generation unit (representing a 20\% share of the capital expenditure \cite{DEA2020GEN}), but not on the distance to shore. \review{Hourly-sampled demand data covering the same ten years used in the siting stage (i.e., 2010 to 2019) is retrieved from \cite{ENTSOEPowerStats}. Then, as in the previous siting stage, time series are resampled at three-hourly resolution by preserving the mean of each consecutive subset of three hours across the entire time horizon.}

\paragraph{Electricity Generation Assets} There are nine technologies available for electricity generation, i.e., offshore and onshore wind, utility-scale and distributed solar PV, run-of-river and reservoir-based hydro, nuclear plants, open- and combined-cycle gas turbines (OCGT and CCGT, respectively). \review{Only a subset of these technologies (i.e., the offshore wind and the gas-fired units) are sized, while installed capacities of onshore wind, solar PV, hydro and nuclear power plants remain fixed throughout the optimisation.} Recall that the technical potentials of the offshore wind sites are inputs from the siting stage. By contrast, those of the remaining generation technologies to be sized in the CEP framework (i.e., OCGT and CCGT) are assumed to be unconstrained. All generation technologies except the gas-fired power plants are assumed to have non-zero installed capacities at the beginning of the optimisation exercise. More specifically, \SI{61.5}{GW} of nuclear power capacity, \SI{33.5}{GW} of run-of-river hydro power capacity and \SI{98.1}{GW} of reservoir-based hydro power capacity are available throughout the selected European countries \cite{JRC_PPDB, JRC_HY}.\footnote{The modelling of run-of-river capacity factors and of inflows into the water storage of reservoir-based plants is detailed in the Supplementary Material.} Existing wind capacity is obtained from \cite{WindDataBase}, where \SI{99.1}{GW} of offshore wind and \SI{160.5}{GW} of onshore wind capacity in various development stages are reported across Europe. Utility-scale solar PV capacity data is retrieved from \cite{SolarDataBase}, where a legacy capacity of \SI{45.5}{GW} is reported throughout Europe. Finally, country-aggregated capacities for distributed PV installations are retrieved from \cite{solarEurope2019}, where the existence of \SI{77.7}{GW} of such installations is reported within European borders.

\paragraph{Electricity Storage Assets} Two technologies are available for storing electricity, namely, pumped-hydro (PHS) and battery storage (Li-Ion). It is assumed that no legacy capacity is available in Europe for the latter. Pumped-hydro units are not sized within the CEP framework at hand and the power ratings of existing plants are retrieved from \cite{JRC_HY}, where a total of \SI{54.5}{GW}/\SI{1950}{GWh} of PHS units are reported.\footnote{The specific durations of these units is estimated on a unit-by-unit basis via a procedure that is detailed in the Supplementary Material.} A summary of the techno-economic data used to instantiate the CEP problem is provided in Table \ref{tab:tech_summary}.

\begin{table}
\centering
\caption{Summary of techno-economic parameters used to instantiate the CEP problem. N/A values denote either i) the lack of a capacity upper bound or ii) economic information which is irrelevant for the purpose of this study. \review{Data sources} are provided in the Supplementary Material.}
\label{tab:tech_summary}
\renewcommand{\arraystretch}{1.2}
\begin{threeparttable}
\begin{tabular}{lcccccc}
\toprule
Technology         & $\underline{\kappa}$  & $\Bar{\kappa}$ & CAPEX                & FOM        & VOM      & Lifetime  \\
                   & GW(h)  & GW(h)    & M\euro/GW(h)              & M\euro/GW(h)-yr & M\euro/GWh & yrs     \\
\midrule
Onshore Wind       & 160.51 & 160.51   & N/A              & 29.47      & 0.00     & 25        \\
Offshore Wind      & 99.14  & $\approx$450.00\tnote{1}      & 1881.08              & 49.11      & 0.00     & 25        \\
Utility-Scale PV   & 45.52  & 45.52    & N/A               & 7.14       & 0.00     & 25        \\
Distributed PV     & 77.69  & 77.69    & N/A               & 5.36       & 0.00     & 25        \\
OCGT               & 0.0    & N/A      & 838.87               & 3.03       & 0.0076   & 30        \\
CCGT               & 0.0    & N/A      & 1005.27              & 7.58       & 0.0053   & 30        \\
Nuclear            & 61.55  & 61.55    & N/A                  & 106.25     & 0.0018   & 40        \\
Run-of-River Hydro & 33.52  & 33.52    & N/A                  & 0.00       & 0.0119   & N/A       \\
Reservoir Hydro    & 98.12  & 98.12    & N/A                  & 0.00       & 0.0152   & N/A       \\
Pumped-Hydro       & 54.54  & 54.54    & N/A                  & 14.20      & 0.0002   & N/A       \\
Battery Storage\tnote{2}             & 0.00   & N/A      & 100.00 / 94.00       & 0.54       & 0.0017   & 10       \\
HVAC               & 100.61 & N/A      & 2.22 & 0.017      & 0.00      & 40        \\
HVDC               & 31.07  & N/A      & 1.76 & 0.021      & 0.00      & 40        \\
\bottomrule
\end{tabular}
    \begin{tablenotes}\footnotesize
    \item[1] Value to be interpreted as a lower bound on the real value which depends on the outcome of the sizing stage (as the potential of each site is proportional to its corresponding surface area which, in turn, depends on the latitude).
    \item[2] For \review{battery storage (Li-ion)}, the values before and after the slash sign under the CAPEX header express the capital expenses for power and energy components. Moreover, the FOM is expressed in M\euro/GW and applies to the power component, while the VOM is expressed in M\euro/GWh and applies to the energy component.
    \end{tablenotes}
\end{threeparttable}
\end{table}

\paragraph{Policy Constraints} A set of policy-related constraints are enforced in the CEP problem. \review{One of the main constraints driving the design of power systems under deep decarbonization targets is the $\mathrm{CO_2}$ budget. In the current exercise, this budget is enforced system-wide and its value imposes a 90\% reduction in carbon dioxide emissions throughout the optimisation horizon relative to 1990 levels. Then, a planning reserve margin of 20\% is considered at each bus $n \in \mathcal{N}_B$ in Eq. (\ref{eq:prm_constraint}). The set $\mathcal{D}$ gathering dispatchable generation units providing firm capacity includes OCGT, CCGT, nuclear and reservoir-based hydro power plants. Furthermore, at each bus $n \in \mathcal{N}_B$, the capacity credit of RES sites is computed during the top 5\% time instants of peak electricity demand.} 

A detailed account of the techno-economic assumptions considered in this study is provided in the Supplementary Material. The input data used to set-up the siting and sizing models is available at \cite{dox_repo}. The code used to run the both models is available at \cite{resite_ip_git} and \cite{replan_git}, respectively.

\section{Results}\label{Results}

In this section, a series of experiments that compare the implications of the proposed siting schemes on the design and economics of power systems is conducted. In particular, the impact of two variants (i.e., partitioned and un-partitioned) of $PROD$ and $COMP$ on the siting of roughly 350 offshore wind power plants in the European power system is discussed.

\subsection{Impact of Siting Decisions on Offshore Production and Residual Load}\label{SitingIntuition}

\begin{figure}
\centering
\begin{subfigure}[b]{0.9\textwidth}
   \includegraphics[width=1\linewidth]{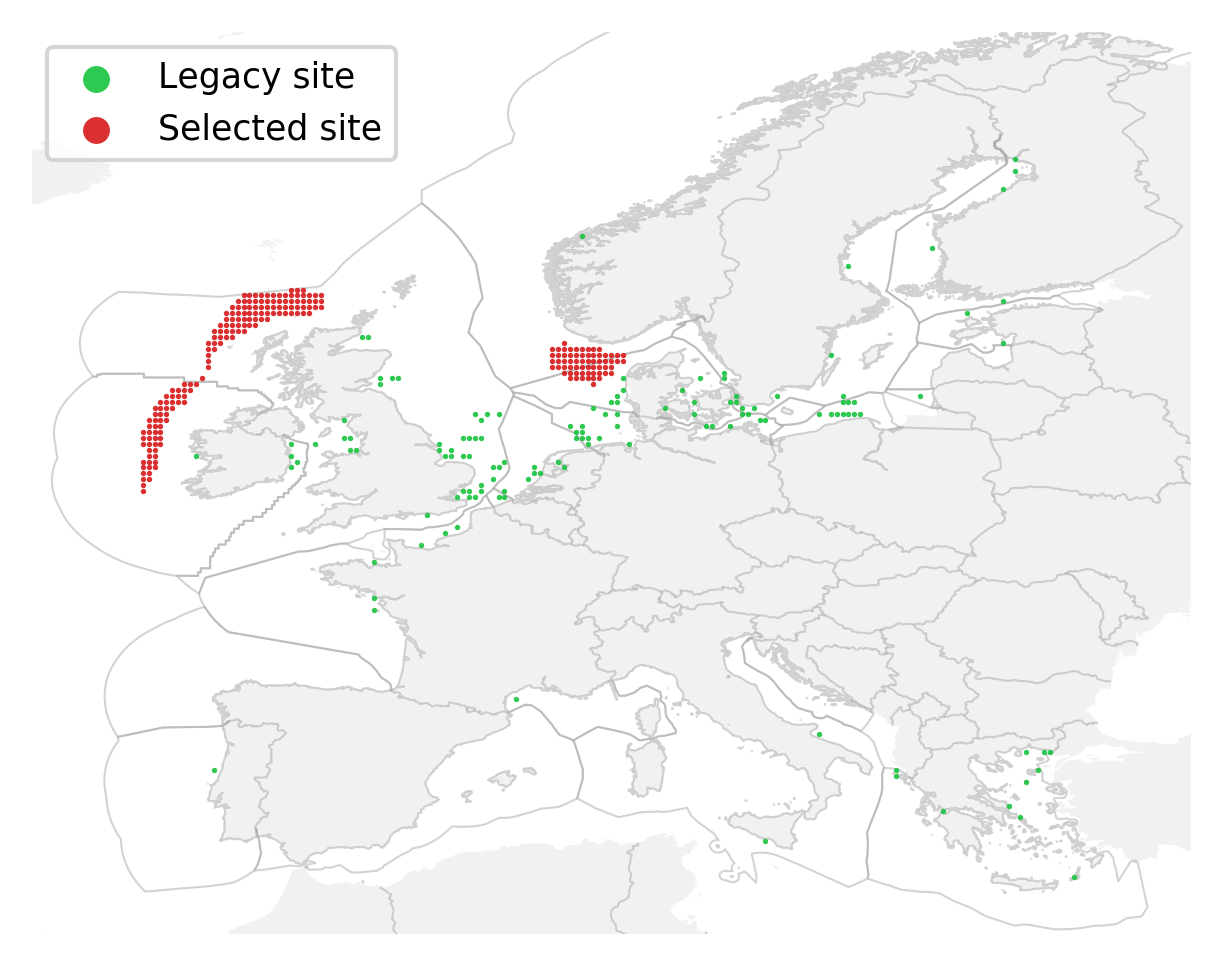}
   \label{fig:k1_PROD} 
\end{subfigure}
\begin{subfigure}[b]{0.9\textwidth}
   \includegraphics[width=1\linewidth]{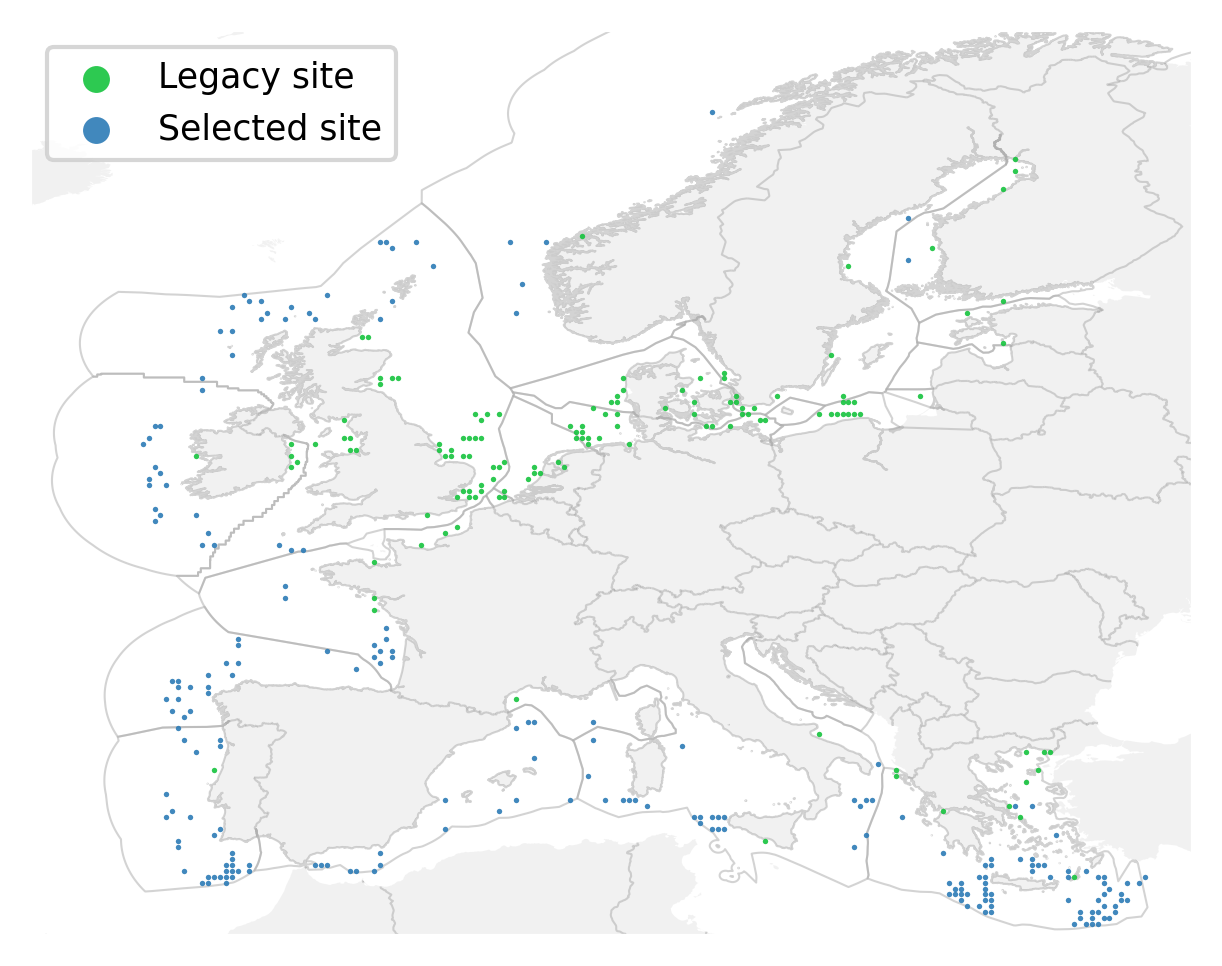}
   \label{fig:k1_COMP}
\end{subfigure}
\caption{Deployment patterns for the $PROD$ (top) and $COMP$ (bottom) siting schemes for the unpartitioned ($B=1$) case. In both plots, legacy locations are displayed in green. Exclusive Economic Areas depicted by the grey contours outside the European land mass.}
\label{fig:k1_deployments}
\end{figure}

\review{The first set of results provides insight into the impact of the two siting strategies on the aggregate offshore wind and residual demand signals. To this end, an unpartitioned set-up is used (i.e., where sites are deployed with no consideration for territorial constraints), whose outcome can be seen in Figure \ref{fig:k1_deployments} for both $PROD$ and $COMP$ schemes, where green markers depict the 135 legacy offshore wind sites. In the top figure, the $PROD$ strategy concentrates all remaining sites to be deployed in two of the most productive areas within the European Seas (i.e., the Atlantic region offshore the British Isles and the North Sea area between Denmark and Norway) \cite{globalwindatlas}. By contrast, the bottom subplot shows that the $COMP$ strategy distributes sites across several distinct areas found within European EEZ. More specifically, offshore wind deployments under this strategy seem to follow two directions. On the one hand, resource-rich sites in the Atlantic region are still exploited, though to a lesser extent considering that the very good resource in the North Sea basin is already well represented in the set of legacy sites. On the other hand, a significant share of the sites picked by $COMP$ are spread in two regions (i.e., Iberia and Southeastern Europe) that are known to have distinct and complementary wind regimes to the ones in Northern Europe, as pointed out in \cite{Grams2017, Cortesi2019}.}

\begin{figure}
\centering
\includegraphics[width=\linewidth]{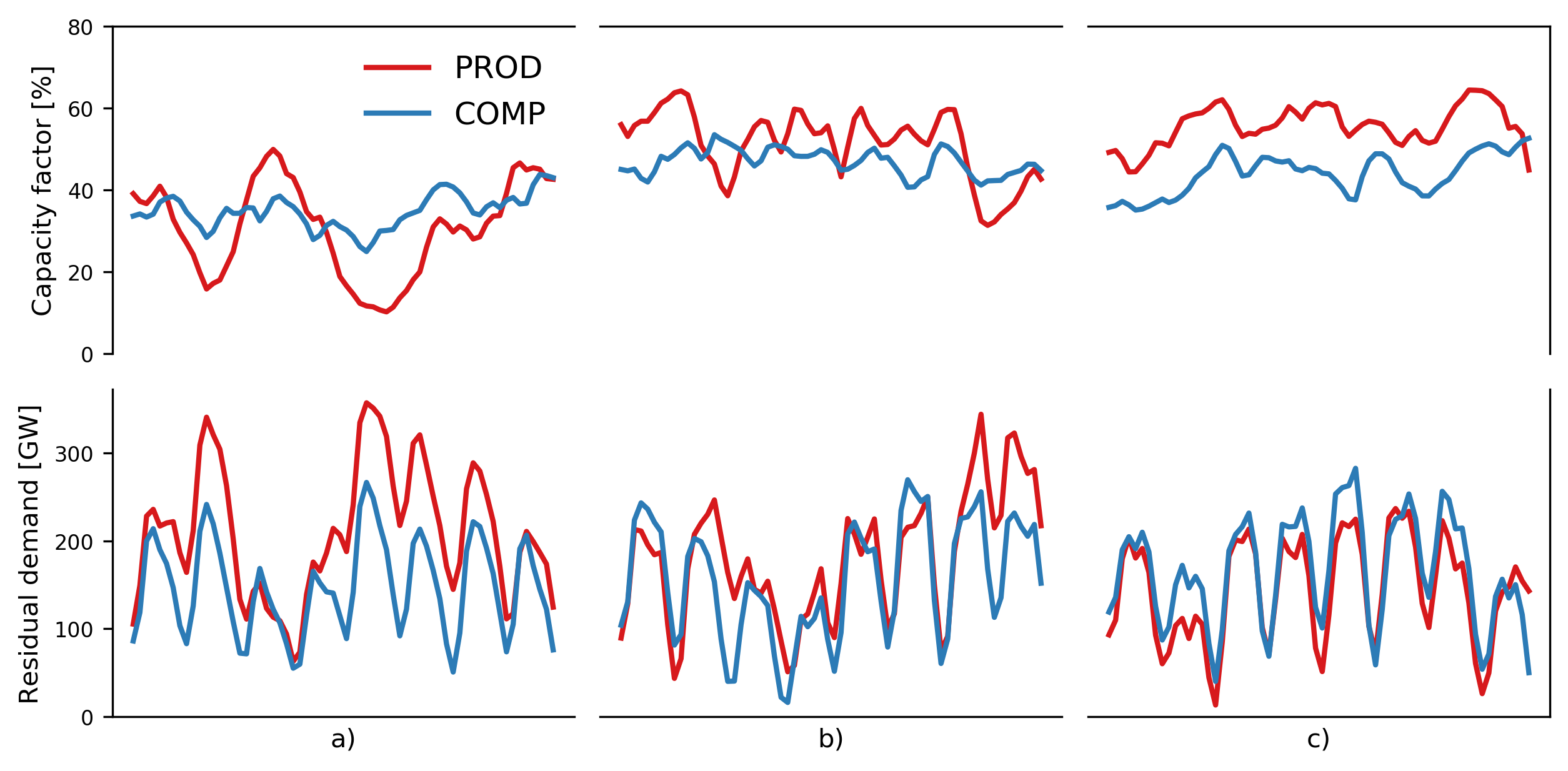}
\caption{Visual examples of aggregate offshore wind (top) and residual demand (bottom) signals for the unpartitioned ($B=1$) $PROD$ and $COMP$ schemes.}
\label{fig:siting_signals_plot}
\end{figure}

\begin{figure}
\centering
\includegraphics[width=\linewidth]{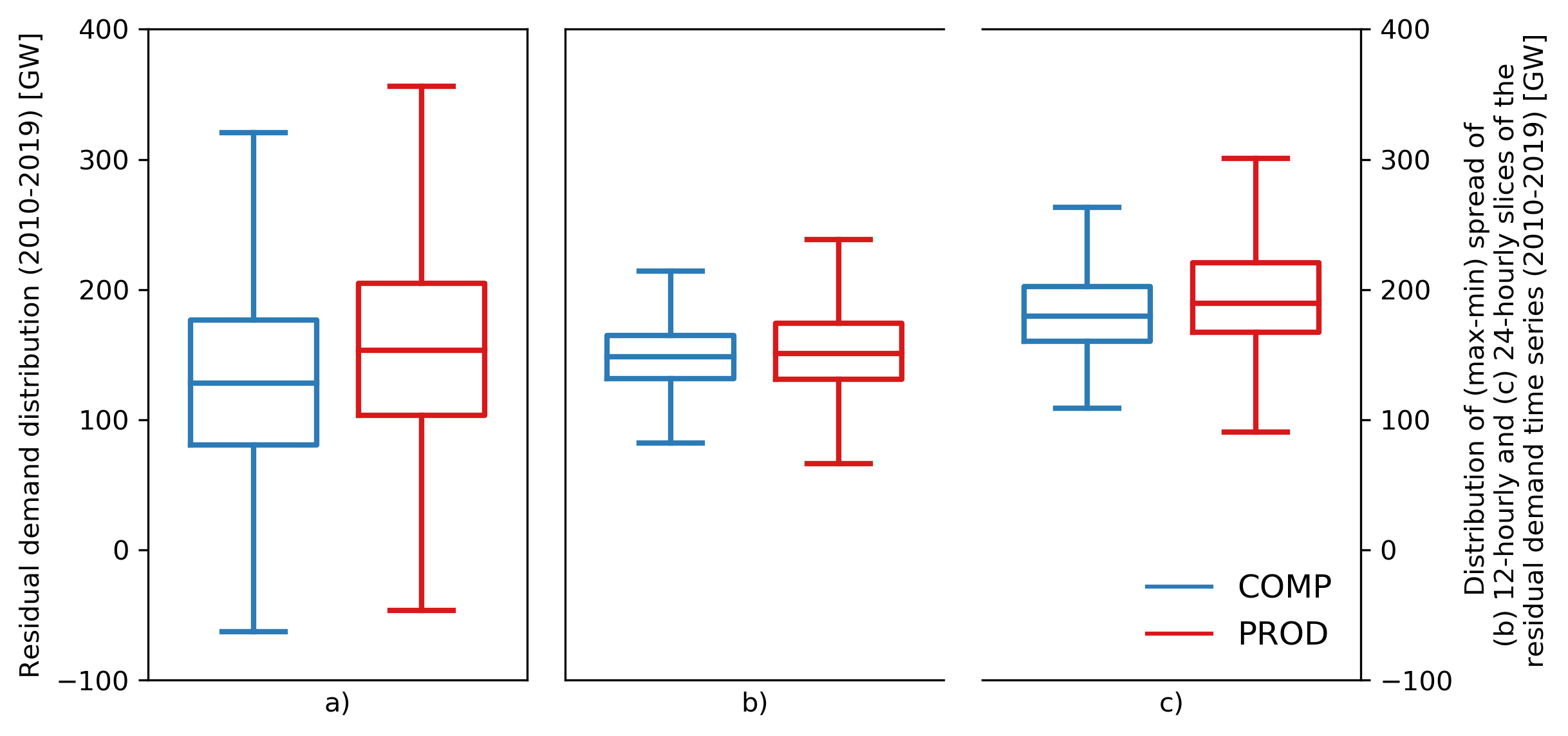}
\caption{Statistical distribution of the residual demand under the unpartitioned ($B=1$) $PROD$ and $COMP$ siting schemes (left). Statistical distribution of the (max-min) spread for 12-hourly and daily disjoint intervals of the residual demand time series under the unpartitioned ($B=1$) $PROD$ and $COMP$ siting schemes (right). Boxes depicting the first quartile, median and third quartile of time series, respectively.}
\label{fig:boxplots_signals_plot}
\end{figure}

\review{The effects of offshore wind power plant siting decisions on the aggregate offshore wind and residual load signals can be seen in Figures \ref{fig:siting_signals_plot} and \ref{fig:boxplots_signals_plot}. More specifically, Figure \ref{fig:siting_signals_plot} displays the aggregate offshore wind signal (top subplot) as well as the aggregate residual demand signal (bottom subplot). The signal shown in the top subplot is obtained by spatially averaging the capacity factor time series of the 353 locations selected by the two siting schemes under consideration, while the aggregate residual demand signal is calculated as follows i) the technology power density and site area assumptions considered in Section \ref{TestCaseSiting} are preserved and ii) the demand signals of every country in Table \ref{tab:site_distributions} are summed to yield a single EU-wide profile from which the aggregate offshore wind feed-in is subtracted. Figure \ref{fig:boxplots_signals_plot}, on the other hand, shows the statistical distribution of the residual demand (\ref{fig:boxplots_signals_plot}a) as well as the statistical distribution of the spread between the maximum and minimum residual demand for 12-hourly and daily (disjoint) time periods (\ref{fig:boxplots_signals_plot}b and \ref{fig:boxplots_signals_plot}c). All distributions are constructed using data from ten weather years (2010-2019).}

\review{Figure \ref{fig:siting_signals_plot} suggests that the $COMP$ scheme is indeed able to select sites with fewer periods of simultaneous low electricity production than the $PROD$. More precisely, aggregate capacity factor values stay between (roughly) 30\% and 60\% for the $COMP$ scheme while the range of capacity factor values covered by the $PROD$ scheme is much broader. This observation is consistent with the fact that the $COMP$ deployment pattern covers 29031 time windows (out of 29218), while the $PROD$ pattern covers 27147 time windows (around 6.5\% fewer than $COMP$), which also implies that instances of high residual load are more frequent in the $PROD$ pattern. This claim is supported by Figure \ref{fig:boxplots_signals_plot}a, which shows that the $COMP$ scheme leads to an overall reduction in residual demand. Indeed, the first quartile, the median, the third quartile and the maximum of the $COMP$ scheme all correspond to significantly lower residual demand values than those of the $PROD$ scheme. Furthermore, Figure \ref{fig:siting_signals_plot} suggests that some degree of aggregate output variability reduction on time scales ranging from hours to days may be obtained as a by-product of the $COMP$ scheme. This intuition is also supported by the box plots in Figures \ref{fig:boxplots_signals_plot}b and \ref{fig:boxplots_signals_plot}c, which indicate that both the full and interquartile ranges of siting patterns produced by the $COMP$ scheme are narrower than those obtained by the $PROD$ scheme. Finally, in Figure \ref{fig:siting_signals_plot}c, it can be seen that the $COMP$ scheme sometimes produces less than $PROD$ for a few days in a row, which can partly be attributed to the fact that the $PROD$ scheme maximises the average capacity factor.}


\begin{figure}
\centering
\begin{subfigure}[b]{0.9\textwidth}
   \includegraphics[width=1\linewidth]{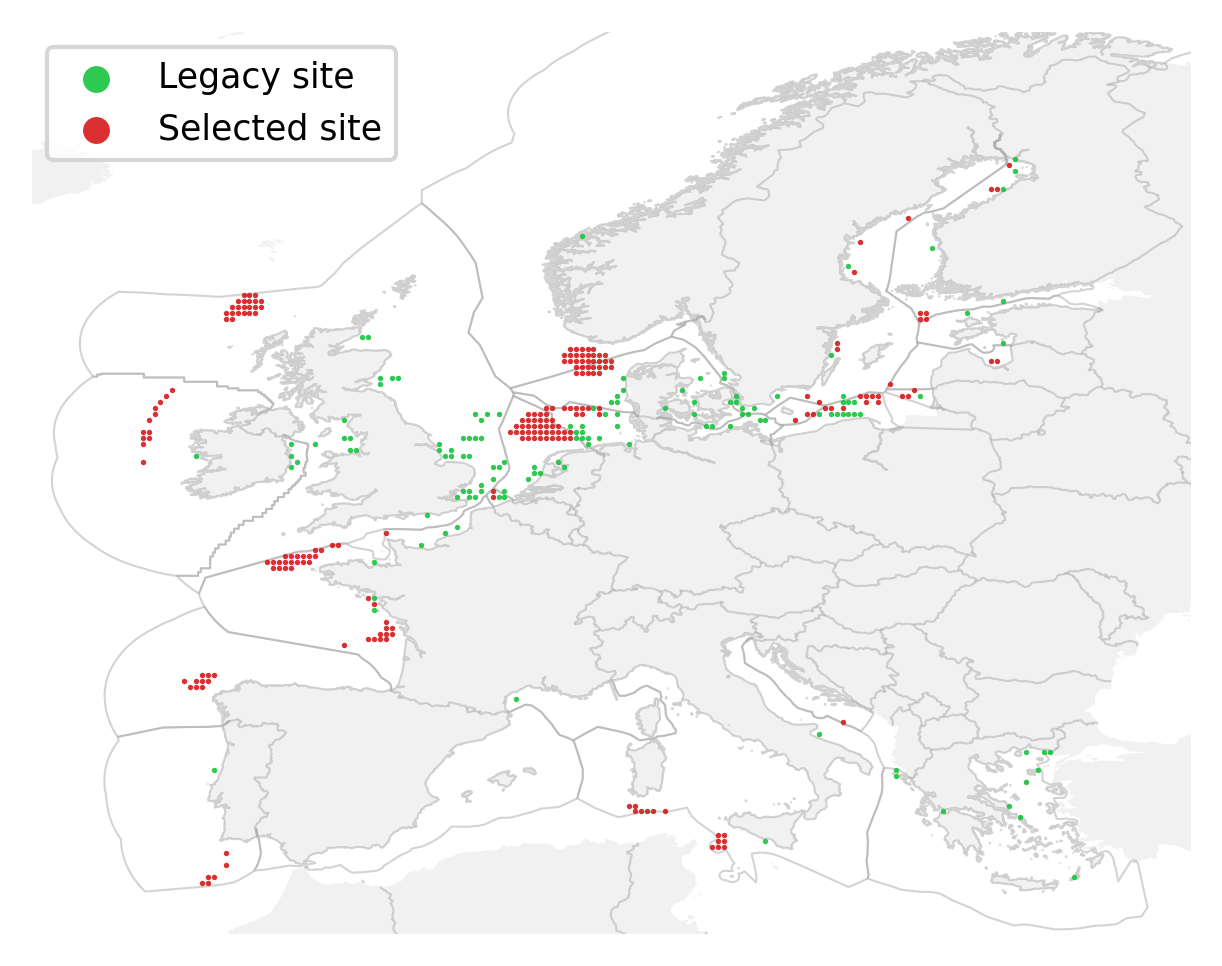}
   \label{fig:k19_PROD} 
\end{subfigure}
\begin{subfigure}[b]{0.9\textwidth}
   \includegraphics[width=1\linewidth]{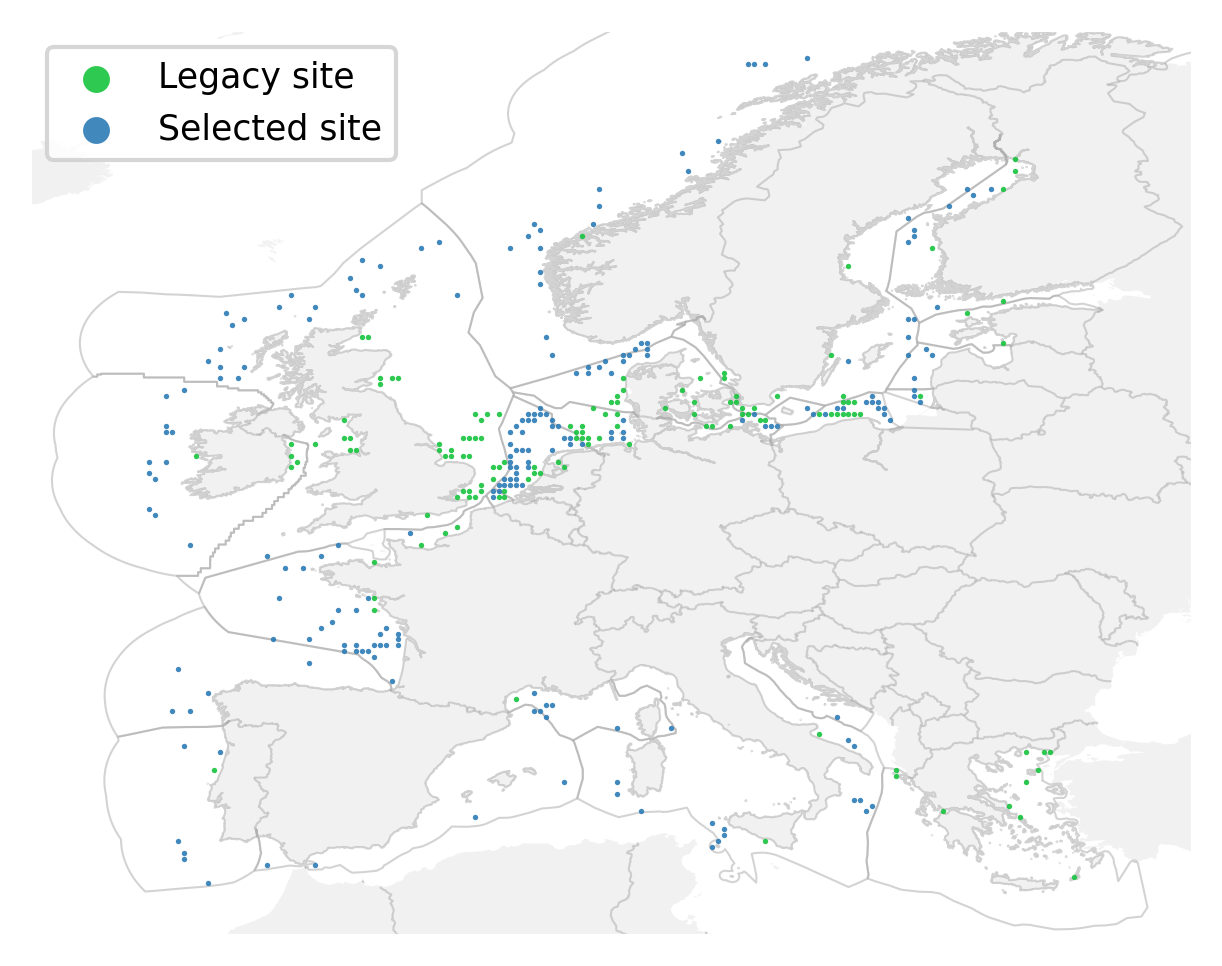}
   \label{fig:k19_COMP}
\end{subfigure}
\caption{Deployment patterns for the $PROD$ (top) and $COMP$ (bottom) siting schemes for the unpartitioned ($B=19$) case. In both plots, legacy locations are displayed in green. Exclusive Economic Areas depicted by the grey contours outside the European land mass.}
\label{fig:k19_deployments}
\end{figure}

\begin{figure}
\centering
\includegraphics[width=\linewidth]{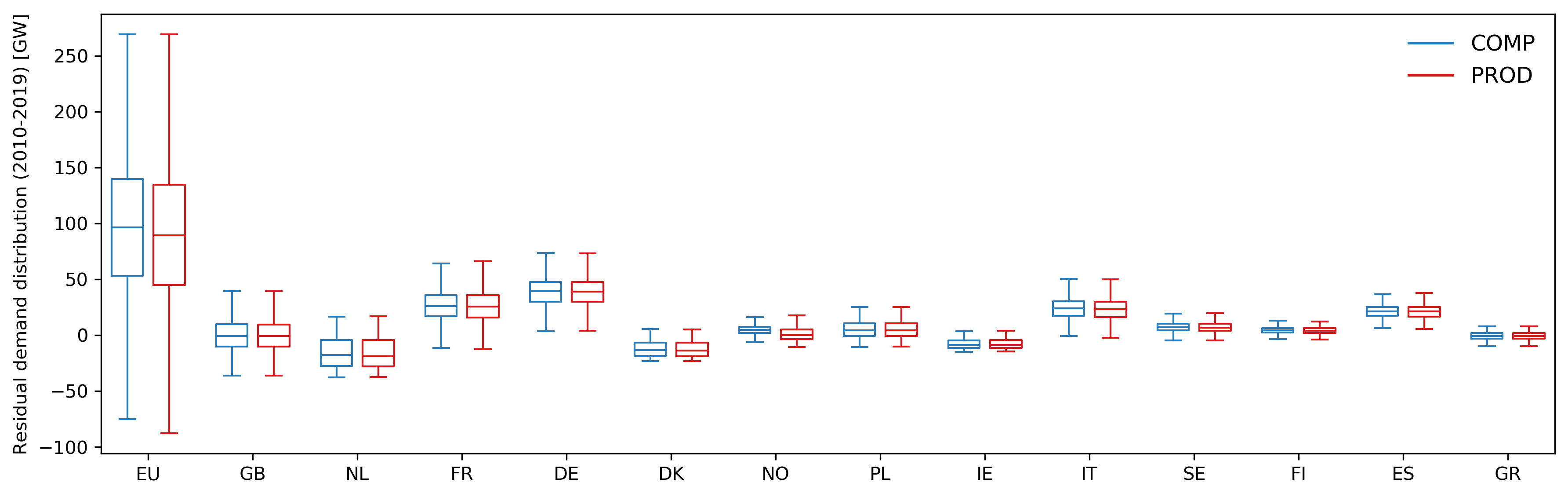}
\caption{Statistical distributions of residual demand time series i) aggregated across Europe and ii) in countries with more than $k_n=10$ deployments under the partitioned ($B=19$) $PROD$ and $COMP$ siting schemes.}
\label{fig:boxplots_signals_plot_k19}
\end{figure}

\review{Variants of the $PROD$ and $COMP$ siting schemes that select locations while satisfying country-based deployment targets ($B = 19$, as shown in Table \ref{tab:site_distributions}) are analyzed next. The associated $PROD$ and $COMP$ deployment patterns are shown in Figure \ref{fig:k19_deployments}, where green markers depict legacy locations. In this context, the $PROD$ scheme (top map) yields a set of clusters of locations, which correspond to the most productive areas of each EEZ. Hence, the resulting deployment pattern is much more scattered than the one observed in the unpartitioned set-up and benefits from much more diverse wind regimes, as suggested by Grams et al. \cite{Grams2017}. The $COMP$ pattern (bottom map) is even more scattered than the $PROD$ one. Legacy locations are common to both schemes and about 19\% of non-legacy locations selected by the $COMP$ scheme are also selected by the $PROD$ scheme, up from 6\% in the unpartitioned set-up. The partitioned $PROD$ and $COMP$ patterns therefore share a total of 176 locations (i.e., roughly 50\% of all selected locations). Furthermore, in several countries, the number of candidate locations available is only slightly greater than the number of locations that must be deployed there. Hence, even though locations selected in these countries by the $PROD$ and $COMP$ schemes may not be exactly identical, they nevertheless end up being in the direct vicinity of one another and therefore experience very similar wind regimes. This is especially true in the Baltic Sea and in countries like Denmark or the Netherlands. This also happens in countries such as France and Ireland, though to a lesser extent, in spite of the fact that the numbers of candidate locations available far exceed the numbers of locations that must be deployed there. Overall, this analysis suggests that the two siting schemes are likely to yield deployment patterns whose performance are comparable. Inspecting the $COMP$ siting objectives achieved by both deployment patterns confirms this intuition. More precisely, the $COMP$ pattern covers 27981 windows, while the $PROD$ pattern covers 27688 windows. In other words, there is only a 1\% difference between them. In addition, Figure \ref{fig:boxplots_signals_plot_k19} shows the distributions of residual demand aggregated across Europe and on a country-by-country basis. At the notable exception of Norway, where the median residual load of the $COMP$ scheme is slightly higher than that of the $PROD$ scheme, the residual demand distributions that both schemes yield are virtually identical. Interestingly, the first quartile, median and third quartile of the EU-wide $COMP$ distribution correspond to residual demand levels that are slightly higher than those observed for the $PROD$ distribution, while maximum residual levels are virtually identical for both schemes. Hence, these results suggest that enforcing country-based deployment targets and selecting locations in the most productive areas is enough to take advantage of the diversity that exists in European offshore wind regimes.}

\subsection{Impact of Siting Decisions on Capacity Expansion Planning Outcomes}\label{SizingResults}

\review{In this section, the impact of different siting schemes on the outcomes of the capacity expansion planning set-up described in Section \ref{SizingModel} are investigated. To this end, the outcomes of the two variants of the two siting schemes introduced in Section \ref{SitingModel} (four in total, i.e., $B=1$ vs $B=19$ for both $PROD$ and $COMP$) are used to run the CEP stage over the ten individual weather years included in the siting optimization problem (i.e., 2010 to 2019).}

\subsubsection{Impact on Power System Economics}

\begin{figure}
\centering
\includegraphics[width=\linewidth]{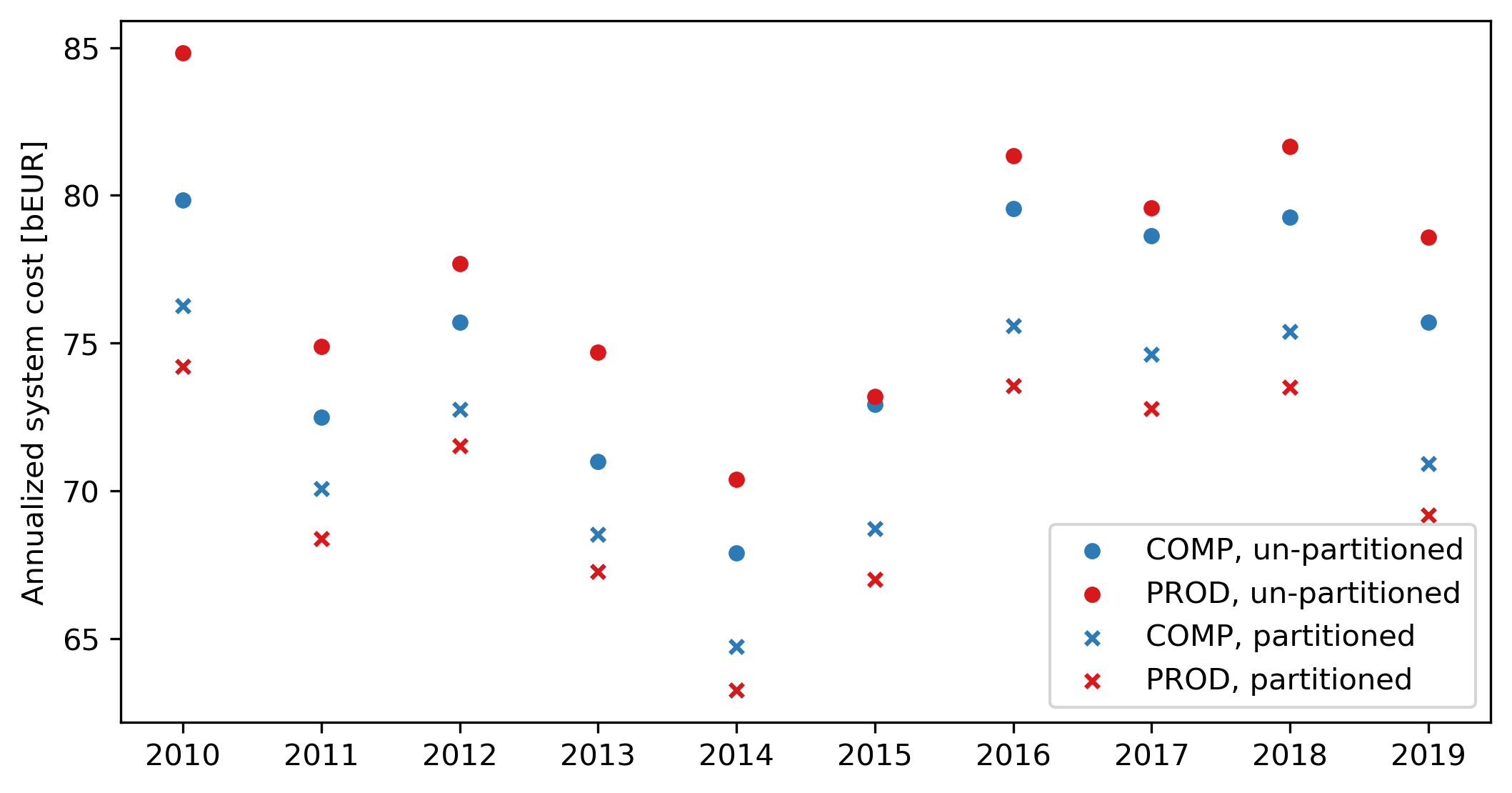}
\caption{Capacity expansion objectives of single-year set-ups instantiated with the outcomes of partitioned ($B=19$) and unpartitioned ($B=1$) $PROD$ and $COMP$ siting schemes.}
\label{fig:single_years_scatter}
\end{figure}

\review{Figure \ref{fig:single_years_scatter} gathers the objectives (i.e., annualized system costs) achieved in the forty aforementioned runs. First, the scatter plot shows distinct trends across partitioned and unpartitioned siting schemes, respectively. On the one hand, $COMP$ seems to outperform $PROD$ consistently (i.e., by 0.4\% to 5.9\% depending on the weather year considered) across the ten weather years when partitioning constraints are not enforced ($B = 1$). By contrast, when country-based deployment targets ($B = 19$) are accounted for in the siting of offshore wind power plants, the $PROD$ scheme leads to annualized system costs that are between 1.2\% and 2.8\% lower than those achieved by the $COMP$ scheme, depending on the weather year considered. In addition, this plot shows a great deal of variability in the sizing objectives achieved by different siting schemes and for different weather years (e.g., differences of up to 20\% between the partitioned and unpartitioned $PROD$ schemes). This suggests that both siting decisions and inter-annual weather variability can have a substantial impact on the economics of power systems relying heavily on weather-dependent renewable generation assets such as offshore wind power plants. Unless otherwise stated, the analyses carried out in the next sections focus on the two extreme weather years (i.e., 2010 and 2014) in order to gain a better understanding of how siting decisions affect power system design.}

\subsubsection{Impact on Power System Design}\label{TableDifferences}

A summary of relevant system design indicators is provided in Table \ref{tab:results_table}, where the CEP outcomes of eight different runs (i.e., two weather years, two siting strategies, two deployment set-ups) are reported.

\review{The first half of Table \ref{tab:results_table} gathers results obtained for CEP instances constructed using unpartitioned deployment patterns ($B = 1$, depicted in Figure \ref{fig:k1_deployments}). A number of observations can be made. First, higher offshore wind capacities are observed in $COMP$-based configurations. This is due to the fact that the average capacity factors (41.5\% and 41.6\% for 2010 and 2014, respectively) are lower than the ones of the set of sites corresponding to $PROD$ (43.0\% and 45.0\% for 2010 and 2014, respectively), an aspect which inherently leads to higher installed capacities in the former scheme (considering that both $PROD$ and $COMP$ are required to meet the same electricity demand profile). More installed capacity leads to more electricity generation from these units in the 2010 instance and also to a significant reduction of curtailment volumes in both weather years considered. Moreover, maybe the most notable effect of deploying sites based on $COMP$ is a reduction in dispatchable capacity requirements. Recall that $COMP$ is designed to minimize the occurrence of system-wide resource scarcity events, such as the one depicted in Figure \ref{fig:siting_signals_plot}a. As a result, the corresponding sizing instances consistently reveal smaller capacities for dispatchable generation units. More specifically, OCGT and CCGT capacities are reduced by up to 18.2\% and 49.8\% compared to the corresponding $PROD$ runs. This also translates into considerably smaller generation volumes from these units, with the exception of OCGT in the 2014 run, where the additional \SI{2.4}{TWh} are used to replace the generation deficit brought up by offshore wind. Furthermore, Li-Ion does not seem to play a significant role in the design of the resulting systems (an aspect that holds across all subsequent runs). This outcome has two main causes. First, considerable pumped-hydro storage capabilities exist as legacy installations in the system under study. Second, the time resolution used in this exercise (i.e., three-hourly) does not capture short-term balancing events for which Li-Ion storage is particularly appealing. With respect to transmission capacities, it appears that $COMP$ is on par with $PROD$ in the 2010 instance (though a 15.7\% increase in flows leads to a more efficient utilization of the infrastructure), while a reduction of 16.4\% is observed in the 2014 run. Put together, these outcomes lead to total system cost reductions under the $COMP$ scheme of 5.9\% (for 2010) and 3.6\% (for 2014). It is worth pointing out that cost savings achieved by a reduction in dispatchable capacity deployment and use are partly offset by an increase in offshore wind capacity deployment, which is comparatively much more expensive per unit capacity.}

\begin{table}
\caption{Comparison of annualized system costs and installed capacities for various technologies sized in the CEP framework. The analysis is conducted for the partitioned ($B=1$) and un-partitioned ($B=19$) variants of the $PROD$ and $COMP$ siting schemes and for a weather year with inferior (i.e., 2010) and superior (i.e., 2014) wind quality, respectively.}
\label{tab:results_table}
\centering
\renewcommand{\arraystretch}{1.2}
\begin{threeparttable}
\begin{tabular}{|ll|cc|cc||cc|cc|}
\toprule
           & \small{Weather Year} & \multicolumn{4}{c||}{2010} & \multicolumn{4}{c|}{2014} \\
           & \small{Siting Scheme} & \multicolumn{2}{c|}{$PROD$} & \multicolumn{2}{c||}{$COMP$} & \multicolumn{2}{c|}{$PROD$} & \multicolumn{2}{c|}{$COMP$}\\
           & \small{Indicator} & $K$\tnote{1} & $p$\tnote{2} & $K$ & $p$ & $K$ & $p$ & $K$ & $p$ \\\midrule
\multirow{7}{*}{\rotatebox[origin=c]{90}{un-partitioned $(B=1)$}} & $\mathrm{W_{off}}$\tnote{3} & 416.4 & 1532.4 & 463.8 & 1679.2 & 397.0 & 1517.9 & 411.3 & 1514.1 \\
&&  & (29.7) &  & (25.4) &  & (41.3) & & (15.7)\\
&OCGT       & 298.6 & 5.3 & 286.2 & 4.7 & 308.6 & 7.1 & 252.4 & 9.5\\
&CCGT       & 73.7 & 179.2 & 36.9 & 56.8 & 25.0 & 24.7 & 20.3 & 21.4\\
&Li-Ion     & 0.01 & N/A & 0.01 & N/A & 0.01 & N/A & 0.01 & N/A \\
&Transm.\tnote{4} & 189.0 & 1802.5 & 188.9 & 2085.3 & 188.8 & 2015.9 & 157.8 & 1839.67\\
&ASC\tnote{5} & \multicolumn{2}{c|}{84.8} & \multicolumn{2}{c||}{79.8} & \multicolumn{2}{c|}{70.4} & \multicolumn{2}{c|}{67.9}\\\midrule\midrule
\multirow{7}{*}{\rotatebox[origin=c]{90}{partitioned $(B=19)$}} & $\mathrm{W_{off}}$ & 464.3 & 1696.3 & 478.8 & 1687.9 & 400.8 & 1521.3 & 412.4 & 1519.4 \\
&&  & (38.4) &  & (39.7) &  & (22.4) & & (23.7)\\
&OCGT       & 268.0 & 7.7 & 267.1 & 7.8 & 230.5 & 9.6 & 226.1 & 9.8\\
&CCGT       & 34.3 & 46.6 & 32.9 & 51.4 & 24.7 & 23.4 & 25.5 & 26.0\\
&Li-Ion     & 0.0 & N/A & 0.0 & N/A & 0.01 & N/A & 0.0 & N/A \\
&Transm. & 124.3 & 1359.3 & 124.4 & 1371.7 & 116.3 & 1319.6 & 115.8 & 1306.9\\
&ASC & \multicolumn{2}{c|}{74.2} & \multicolumn{2}{c||}{76.3} & \multicolumn{2}{c|}{63.3} & \multicolumn{2}{c|}{64.7}\\\bottomrule
\end{tabular}
    \begin{tablenotes}\footnotesize
    \item[1] $K$ denotes the system-wide capacity of a given technology (incl. legacy capacity), as resulted from the optimisation exercise and it is expressed in energy units. For instance, capacities of generation technologies (e.g., offshore wind, OCGT, CCGT) are reported in GW. For lithium-ion storage (Li-Ion), the same quantity it is expressed in GWh, while transmission capacities are expressed in TWkm.
    \item[2] $p$ denotes the amount of electricity produced (for generation technologies) or transported (for transmission technologies) across over a full year. Values are expressed in TWh.
    \item[3] Values in parentheses represent offshore wind curtailment volumes (expressed in the same units as $p$).
    \item[4] In this table, both electricity transmission technologies (i.e., AC and DC) are aggregated into one term.
    \item[5] ASC stands for "annualized (total) system cost", is expressed in billion \euro \ and represents the objective function of the expansion planning stage.
    \end{tablenotes}
\end{threeparttable}
\end{table}

\review{Results pertaining to CEP instances constructed from partitioned deployment patterns ($B = 19$, shown in Figure \ref{fig:k19_deployments}) are provided in the second half of Table \ref{tab:results_table}. To begin with, the number of non-critical windows obtained for these set-ups reveal a much tighter difference between the two deployment schemes, i.e., 94.71\% (or 27688) and 95.77\% (or 27981) for $PROD$ and $COMP$, respectively. In terms of system costs, partitioned $PROD$ regularly out-performs partitioned $COMP$, as already revealed in Figure \ref{fig:single_years_scatter}. Specifically, the latter scheme leading to system configurations which are 2.8\% (for the 2010 instance) and 2.3\% (for the 2014 case) more expensive than the corresponding $PROD$-based runs. This outcome can be explained as follows. Regardless of the weather year considered, the $COMP$-based runs deploy more offshore wind capacity (additional 3.1\% and 2.9\% in the 2010 and 2014 runs, respectively), which translates into higher capital expenditures. Nevertheless, the associated generation levels are slightly inferior to those of the $PROD$-specific instances due to the differences between the capacity factors of the sets of sites associated with the two siting schemes (recall that $PROD$ is by design selecting the locations with the highest capacity factors in all $B=19$ regions). More specifically, the average capacity factors for the $PROD$ set of $k=353$ sites are 42.3\% and 43.8\% (2010 and 2014, respectively), compared to 40.9\% (during 2010) and 42.3\% (during 2014) for the $COMP$ set of locations. This time, however, the $PROD$ deployment pattern also exploits a great deal of resource diversity itself, which leads to significantly mitigated dispatchable capacity and generation requirements compared to the un-partitioned $PROD$ case. The $COMP$ siting scheme enables, even in these conditions, an overall capacity reduction of dispatchable units (i.e., of \SI{2.4}{GW} and \SI{3.6}{GW} in 2010 and 2014, respectively) compared to $PROD$, which indicates that the complmentarity-based siting method still manages to provide a set of sites that decreases the peak residual demand across the system. However, power generation from VOM-intensive dispatchable power plants is now used to cover for the offshore wind feed-in deficit, thus resulting in increased O\&M expenditures compounding the additional capital costs due to wind offshore deployments. Finally, no significant differences can be seen in terms of transmission capacity or transited volumes. These results suggest that, as opposed to the un-partitioned scenario, $COMP$ slightly under-performs compared to a siting strategy that assumes the deployment of the most productive offshore sites across the 19 Exclusive Economic Zones.}

\subsection{Sensitivity Analysis}

\review{In this section, a sensitivity analysis is performed in order to evaluate the robustness of results obtained for the partitioned deployment patterns with respect to offshore wind cost assumptions and inter-annual weather variability.}

\subsubsection{Impact of Offshore Wind Cost Assumptions}

\begin{figure}
\centering
\includegraphics[width=\linewidth]{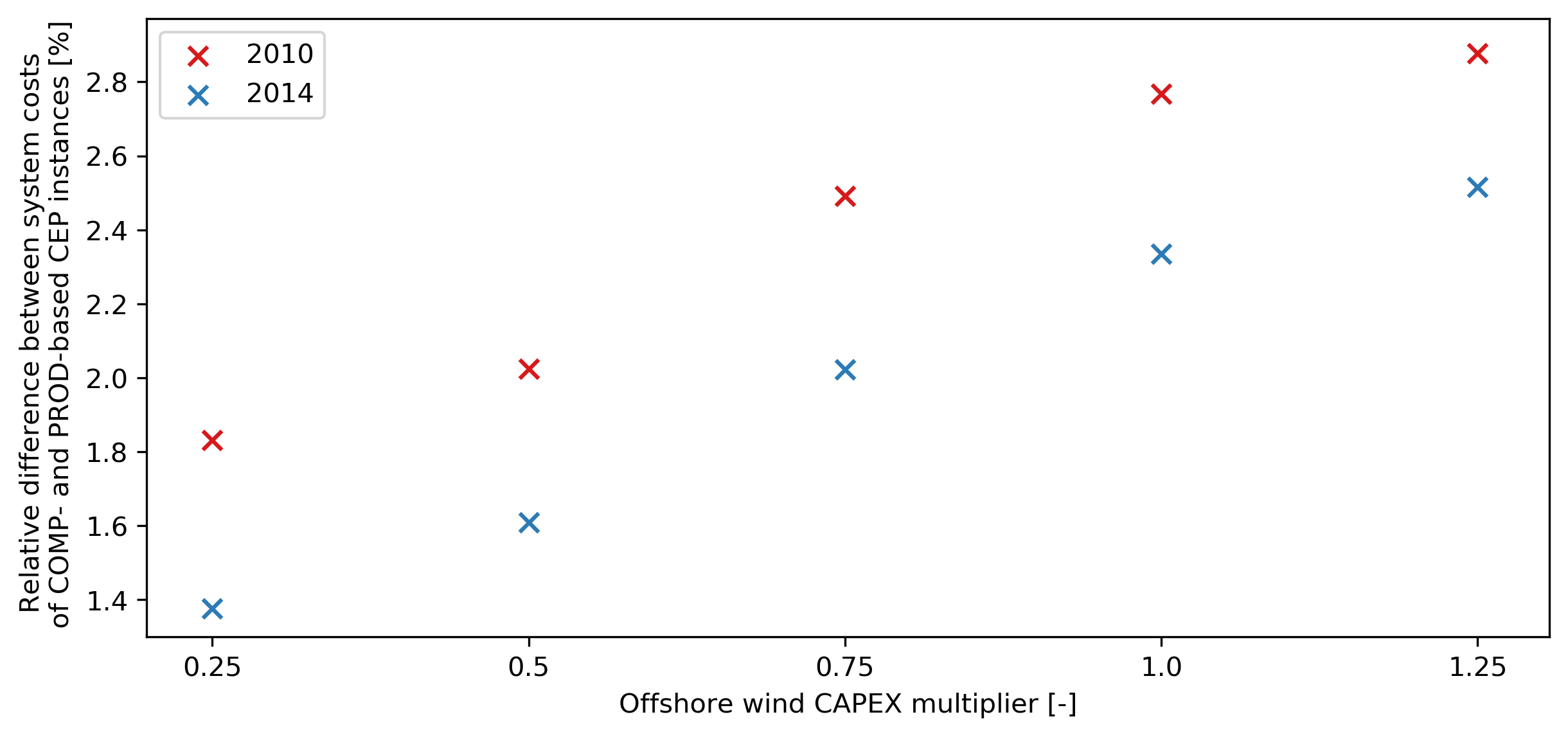}
\caption{Relative differences in annualized system costs achieved by expansion planning instances using the outcomes of partitioned ($B=19$) $COMP$ and $PROD$ siting schemes, for different offshore wind CAPEX multiplicative factors. The analysis is carried out for two weather years (i.e., 2010 and 2014).}
\label{fig:ocm_scatter}
\end{figure}

\review{In view of recent offshore wind cost projections suggesting that costs are likely to decrease substantially by 2050 \cite{Wiser2021} and considering the small difference between annualized system costs reported in Table \ref{tab:results_table} for the $PROD$ and $COMP$ schemes (especially for the partitioned set-ups), evaluating the sensitivity of these results to the economic assumptions laid out in Table \ref{tab:tech_summary} is warranted. More precisely, the outcomes of the partitioned $PROD$ and $COMP$ schemes are used in CEP set-ups where the capital expenditure of offshore wind is varied between 25\% and 125\% of the reference cost, by increments of 25\%. The results of this experiment are gathered in Figure \ref{fig:ocm_scatter}, where the red and blue markers represent the relative difference (in percentage points) between the objectives of $PROD$- and $COMP$-based runs (a positive value indicates higher costs for the latter) for the 2010 and 2014 weather years, respectively.}

\review{It is clear from Figure \ref{fig:ocm_scatter} that $COMP$-based power system designs are consistently more expensive than their $PROD$ counterparts, regardless of the offshore wind cost. The relative difference between the sizing objectives decreases steadily as the value of the cost multiplier decreases and falls below 2\% for both weather years considered, when offshore wind CAPEX is assumed to be only 25\% of the reference value. Overall, the $COMP$ schemes lead to system designs that are 1.37\% to 2.51\% and 1.83\% to 2.88\% more expensive than their $PROD$ counterparts for the 2014 and 2010 weather years, respectively. The main reason behind $PROD$ consistently leading to cheaper system configurations is the fact that, regardless of the offshore wind cost, the total cost of $COMP$-based system designs is offset by additional offshore wind capacity deployments. As already discussed in Section \ref{TableDifferences}, this outcome is driven by lower average capacity factors for the $COMP$ sites compared to the ones corresponding to the $PROD$ sets of locations.}

\subsubsection{Impact of Inter-Annual Weather Variability}\label{Results10yruns}

\review{The variance in the sizing outcomes obtained for problem instances with a time horizon of one year (shown in Figure \ref{fig:single_years_scatter}) supports previous findings suggesting that inter-annual weather variability may have a substantial impact on the cost of operating power systems with high shares of RES-based generation \cite{Collins2018}. Consequently, this experiment seeks to evaluate the performance of power system designs obtained with the partitioned $PROD$ and $COMP$ schemes when the full time series of weather data leveraged in the siting stage (i.e., 2010-2019) is used to instantiate the CEP problem that sizes the system.}

\begin{table}
\centering
\caption{Breakdown of installed capacities and costs per technology for the partitioned $PROD$- and $COMP$-based system designs obtained for a sizing problem instance with a time horizon of ten years (corresponding to the ten weather years used in the siting stage, namely 2010-2019).}
\label{tab:10yruns}
\begin{tabular}{ccrcccc}
\toprule
Scheme                & ASC               &         & OCGT & CCGT & $\mathrm{W_{off}}$ & Transm.  \\\midrule
\multirow{4}{*}{\shortstack[c]{$PROD$}}  & \multirow{4}{*}{\shortstack[c]{722.68 \\ (b\euro)}} & $K$ (GW/TWkm) &   291.3 & 36.4 & 435.7 & 129.4 \\
                        &                   & CAPEX (b\euro)   &   106.6   & 17.4 & 420.6 & 41.7 \\
                        &                & $p$ (TWh) & 88.0 & 590.6 & 16512.5 & 14270.5 \\
                        &                   & OPEX (b\euro) & 8.2 & 41.1 & 0.0 & 0.0 \\\midrule
\multirow{4}{*}{\shortstack[c]{$COMP$}}  & \multirow{4}{*}{\shortstack[c]{740.22 \\ (b\euro)}} & $K$ (GW/TWkm) & 288.4 & 36.6 & 444.3 & 129.4 \\
                        &                   & CAPEX (b\euro)   & 105.6 & 17.5 & 431.4 & 41.7 \\
                        &                & $p$ (TWh) & 80.4 & 696.3 & 16334.0 & 14147.4\\
                        &                   & OPEX (b\euro)   & 7.5 & 48.4 & 0.0 & 0.0 \\\bottomrule
\end{tabular}
\end{table}

\review{The figures in Table \ref{tab:10yruns} indicate that the intuition provided by the sizing runs relying on extreme weather years (see Section \ref{TableDifferences}) still holds when the inter-annual variability of the offshore wind resource is properly accounted for. More specifically, the system configuration based on the $COMP$ siting strategy is 2.4\% more expensive than the one relying on the $PROD$ deployment scheme. Two main factors are behind this cost difference. First, the partitioned $PROD$ siting scheme naturally yields a collection of very productive offshore wind sites. Indeed, a ten-year average capacity factor of 43.2\% is achieved across the 353 sites. Moreover, as previously reported in Section \ref{TableDifferences}, the partitioned $PROD$ deployment pattern achieves a $COMP$ siting objective that is only 1\% lower than that of the $COMP$ deployment pattern. On the other hand, the average capacity factor of the 353 $COMP$ sites is around 41.9\%. This drives the investment in an additional \SI{8.6}{GW} of offshore wind capacity in the $COMP$ system, which represents more than half of the annualized system cost difference between $PROD$ and $COMP$ in Table \ref{tab:10yruns}. Second, although the $COMP$ siting scheme leads to a system design with more offshore wind capacity, the slightly inferior capacity factors lead to a generation deficit of \SI{178}{TWh} across the ten-year optimisation horizon. Legacy generation units, e.g., run-of-river or reservoir-based hydro plants, with non-zero operating costs (as opposed to offshore wind generation) and CCGT power plants (with high O\&M costs) are used to cover the aforementioned shortfall. In total, the additional operating costs incurred by this shift from offshore wind to other generation technologies make up for the remainder of the total cost difference observed between the $PROD$- and $COMP$-based configurations in Table \ref{tab:10yruns}.}


\subsection{Discussion}


\review{It is worth pointing out that, even in the most extreme of situations (represented in Figure \ref{fig:single_years_scatter} by the 2010 un-partitioned set-up), the relative cost difference between system designs using $PROD$ or $COMP$ siting outcomes does not exceed 6\%. The reasons for this limited difference are twofold. First, the case study proposed in this paper investigates solely the siting of offshore wind sites. In the ten-year runs detailed in Section \ref{Results10yruns}, offshore wind represents 38.7\% and 45.4\% of the Europe-wide total installed capacity and generation volumes, respectively (values for the $COMP$-based run), while the remainder corresponds to power generation, storage and transmission technologies which are modelled in an identical fashion in both $PROD$- and $COMP$-based CEP set-ups.
Therefore, the system cost differences identified throughout Section \ref{SizingResults} should be interpreted accordingly as the economic impact of one offshore wind siting strategy or the other on the design of the power system. Second, it should be emphasized that 135 out of a total of $k=353$ sited offshore locations (i.e., a share of 38\%) belong to the subset of legacy sites. This aspect further explains the relatively limited differences in total system costs as, in practice, only 218 offshore wind locations are being sited via the investigated siting strategies. In other words, at most 218 offshore wind resource profiles can differ between the outcomes of $COMP$ and $PROD$.}


\review{An observation consistently made throughout the results section is that $COMP$ outperforms $PROD$ as long as the latter does not fully exploit the resource diversity available across European EEZs (i.e., the un-partitioned set-ups). This suggests that concentrating offshore wind installations in rich, but relatively limited geographical scopes (e.g., the North Sea \cite{WindPowerHub2021}), while deferring their deployment in regions swept by distinct wind regimes (e.g., the Baltic or Mediterranean areas) could lead to undesirable outcomes. One example of such an outcome would be the heavy deployment of thermal dispatchable capacity that would be required to guarantee system adequacy which, considering the duration of investment cycles in the power sector and unless carbon-neutral fuels can be used, would lead to investment decisions that are not consistent with the pathways enabling the achievement of ambitious climate targets by 2050 \cite{IEA2050}}.

\review{Another finding in Section \ref{TableDifferences} concerns the differences observed between the system configurations leveraging the un-partitioned and partitioned siting schemes, respectively. In particular, the two $COMP$ schemes will be discussed. On the one hand, as the un-partitioned set-up is a relaxation of the partitioned case, the former outperforms the latter in terms of siting scores. Specifically, the $B=1$ case yields a set of locations covering 99.36\% of the time windows, while 95.77\% of the time windows are non-critical under the $B=19$ set-up. Interestingly, the superior siting score of the unpartitioned scheme does not translate into a cheaper system configuration, as observed in Figure \ref{fig:single_years_scatter}, where \textbf{x} markers fall below the \textbf{o} markers, regardless of the weather year considered. }

\review{This outcome can be partly explained by the workings of the system adequacy constraint (\ref{eq:prm_constraint}) of the CEP framework, according to which offshore wind (as any other RES technology) can contribute to the provision of firm capacity. More precisely, this constraint is such that system adequacy must be ensured at country-level in order to avoid situations where certain countries excessively depend on electricity imports. Thus, in the partitioned set-up, offshore wind contributes to the provision of firm capacity in all $B=19$ countries across Europe where capacity could be deployed \cite{WindEurope2020_main}. On the contrary, ignoring the partitioning constraints (i.e., $B=1$) results in some countries having less (e.g., Germany, the Netherlands, Norway, etc.) wind deployments compared with the partitioned set-up, as seen in Figure \ref{fig:k1_deployments}. For those countries, the offshore wind potential (which is proportional to the number of deployed sites) becomes lower than in the partitioned set-up (where more sites were deployed) and, in turn, cannot contribute as much to system adequacy. The two dispatchable power generation technologies sized in the CEP problem (i.e., OCGT and CCGT) become the alternatives for firm capacity (since other RES are not sized in the CEP) and the optimiser ends up deploying additional OCGT capacity due to its lower cost per unit capacity, thus augmenting the total system cost of the unpartitioned set-up.}

\review{Another explanation for the partitioned cases outperforming the un-partitioned ones in terms of system cost pertains to the limited amount of system-related information that is made available to the siting stage, irrespective of the siting strategy considered. An example of such information whose implications are relatively easy to gauge are the network constraints. Recall that the siting stage relies solely on renewable resource and electricity demand data and that the classification of time windows is oblivious to limits on transmission capacity between regions. In consequence, even though the unpartitioned $COMP$ siting scheme leads to a higher siting score than the partitioned case, the CEP stage does not manage to take full advantage of the offshore sites identified in the $B=1$ set-up. Indeed, on average (across the ten single-year $COMP$ runs reported in Figure \ref{fig:single_years_scatter}), the capacities deployed at 92.2\% of the 218 offshore sites selected by the partitioned $COMP$ scheme (excluding legacy sites) exceed \SI{100}{MW}, while only 73.4\% of them exceed the same capacity threshold in the CEP set-up based on the unpartitioned $COMP$ outcome. Most of the unexploited sites in the $B=1$ scheme are located in the densely deployed areas of Southwestern and Southeastern Europe. From a siting perspective, the wind regimes of these regions are particularly appealing, since they differ from the ones that prevail in the Northern half of the continent \cite{Grams2017}. Nevertheless, in this case, the benefits of resource diversity cannot be reaped due to the limited options for electricity transmission between the Iberian peninsula and Central-Western Europe, and between Greece and Central Europe.}

\section{Conclusion}\label{Conclusion}

In this paper, a realistic case study evaluating the role that offshore wind power plants may play in the European power system is proposed, with a particular focus on the impact that plant siting strategies have on system design and economics. The paper builds upon a method that combines a siting stage selecting a subset of promising locations for deployment and a capacity expansion framework identifying the power system design that supplies pre-specified demand levels at minimum cost while satisfying technical and policy constraints. In the interest of transparency, an open source tool implementing the two-stage method is also made available.

Two types of deployment schemes that select sites so as to maximise their aggregate power output ($PROD$) and spatiotemporal complementarity ($COMP$) are analysed. Two variants of these siting schemes are also considered, wherein the number of sites to be selected is specified on a country-by-country basis rather than Europe-wide. A few hundred sites are identified by each scheme using a high resolution grid and ten years of reanalysis data, and these sites are then passed to a capacity expansion planning framework in order to assess the impact of siting decisions on power system design and economics. The framework relies on a stylised model of the European power system where each country corresponds to an electrical bus and includes an array of power generation and storage technologies. The framework seeks to size gas-fired power plants, offshore wind power plants, battery storage and electricity transmission assets and operate the system in order to supply electricity demand levels consistent with current European electricity consumption at minimum cost while reducing carbon dioxide emissions from the power sector by 90\% compared with 1990 levels and taking a broad range of legacy assets into account. A detailed sensitivity analysis is also performed in order to evaluate the impact of offshore wind cost assumptions and inter-annual weather variability on system design and economics.

Results show that the $COMP$ scheme yields deployment patterns that have both a much steadier aggregate power output and much lower residual load levels than the $PROD$ scheme if sites are selected without enforcing country-based deployment targets. However, when such constraints are enforced, the siting schemes produce deployment patterns that lead to similar levels of residual load. This suggests that systematically deploying offshore wind sites in the most productive areas of most European countries makes it possible to take full advantage of the diverse wind regimes available in European seas. In addition, power system designs obtained using $COMP$ deployment patterns consistently feature more offshore wind capacity and less dispatchable capacity than $PROD$-based designs. This difference does not always translate into power systems that are cheaper for either of the siting schemes. More precisely, the $COMP$ scheme leads to system designs that are up to 5\% cheaper than $PROD$-based ones when sites are selected without enforcing country-based deployment targets. When such targets are enforced, however, the $PROD$ scheme leads to system designs that are consistently 2\% cheaper than $COMP$-based ones. These results are shown to hold under a broad range of offshore wind cost assumptions and are not affected by inter-annual weather variability.

In future work, several directions can be envisioned for refining the analysis. First, integrating the siting of other RES technologies (e.g., onshore wind, solar PV) into the proposed two-stage method would be of interest to evaluate their synergies in supplying European demand at minimal cost. Then, enhancing the network modelling by i) using a higher spatial resolution and a refined topology, ii) relying on a better approximation of network flows (e.g., a DC-OPF model) would improve the accuracy of the analysis. Evaluating the impact of unit commitment costs and constraints on system designs obtained for different siting schemes would also be of interest. Finally, representing the effect of short-term RES uncertainty in dispatch decisions could also provide some insight into the benefits that different siting schemes may bring about.


\clearpage
\section*{Appendix}
\section*{A1. Siting}
\subsection*{Candidate sites}

\begin{figure}[h]
\centering
\includegraphics[width=\linewidth]{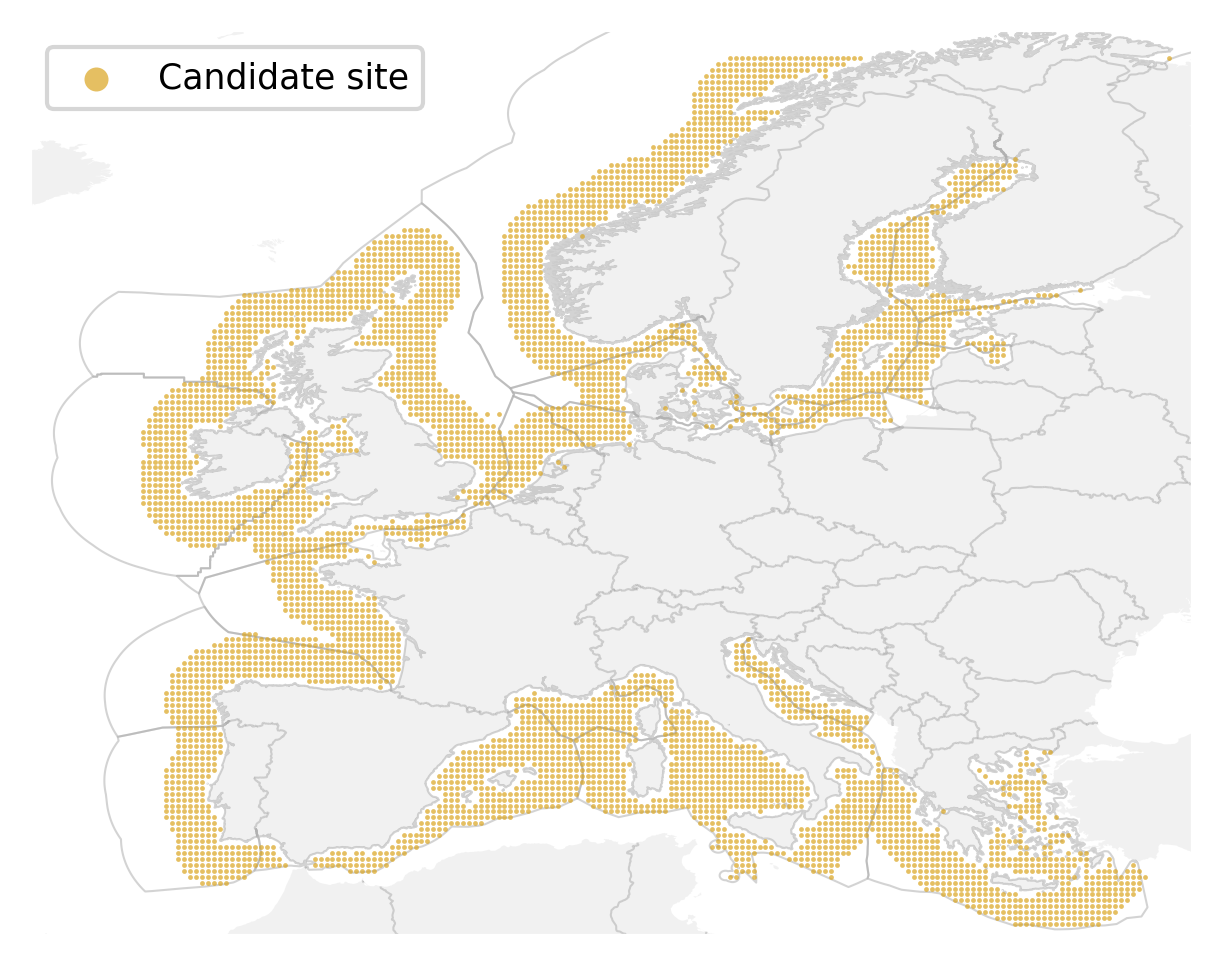}
\caption{Set of candidate locations for offshore wind deployment (in yellow).}
\label{fig:init_coords}
\end{figure}

The set of candidate sites for offshore wind deployment is shown in Figure \ref{fig:init_coords}. In this plot, it can be easily seen that the latitude and distance-to-shore filters are the ones mainly driving the selection of candidate sites. In addition, candidate sites within several Exclusive Economic Zones (EEZ) are not included in this study for different reasons. First, as Albania, Bosnia and Herzegovina, as well as Montenegro are not considered in this study, their EEZs and the corresponding candidate offshore wind sites are not shown in Figure \ref{fig:init_coords}. Second, the EEZ of the Russian exclave of Kaliningrad is not considered for similar reasons. In addition, the territorial waters around the British dependencies of Jersey and Guersney are not included, as they are not officially part of the EEZ of Great Britain.

\section*{A2. Sizing}
\subsection*{Network topology}

\begin{figure}
\centering
\includegraphics[width=\linewidth]{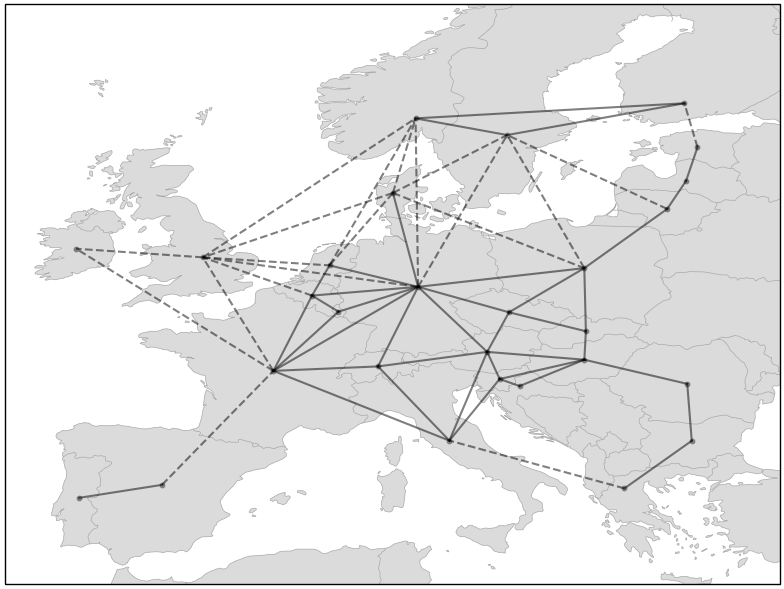}
\caption{Network topology used in the capacity expansion planning stage.}
\label{fig:topology}
\end{figure}

The power system topology used in the capacity expansion planning stage is based on the 2018 version of the Ten Year Network Development Plan of ENTSO-E \cite{TYNDP2018}. More specifically, each country is represented by one node (to which demand and generation signals are attached) and a copper plate assumption is made for power flows within its borders. The links between countries represent aggregations of the total cross-border capacities projected in 2027. For costing purposes, all interconnectors crossing bodies of water are assumed to be developed as DC cables, while the remainder are assumed to be built as AC underground cables. These assumptions can be visualized in Figure \ref{fig:topology}.

\subsection*{Hydro Modelling}\label{HydroModelling}

Hydro power plants modelling requires specific attention, as their operation is inherently constrained by resource- and design-related aspects, e.g., reservoir size, water inflows. In modelling the run-of-river (i.e., ROR) units, inspiration is drawn from the operation of variable RES technologies (e.g., wind, solar PV). Indeed, these former class of plants relies on the availability of water flows to operate, as they do not have significant storage capabilities to regulate their operation. Therefore, after distributing the \SI{33.5}{GW} of ROR capacity \cite{JRC_HY} across the 28 countries (based on their geo-location), these plants are modelled as non-dispatchable, aggregated installations whose per-unit, hourly capacity factors are assumed proportional to the runoff\footnote{Surface and sub-surface water draining away from precipitation, snow melting, etc. Hourly runoff time series covering the studied temporal horizon (2010 to 2019) are obtained from the same reanalysis dataset as wind speeds and solar irradiation data \cite{ERA5}.} available in the country where the aggregate ROR capacity is located. 

More specifically, let $\mathcal{C}$ be the set of all countries and $\mathcal{T}$ be the set of all time instances under study. For any country $c \in \mathcal{C}$, the runoff time series of all reanalysis data points found within its borders are first summed up (\ref{eq:ror_ro_sum}) and then normalized (\ref{eq:ror_ro_norm}). Subsequently, data outliers (i.e., in this case, flood-related events during which the runoff spikes for short periods of time) are removed via an additional step in which a country-specific and dimensionless \textit{flood event threshold} is used to clip such entries, as shown in (\ref{eq:ror_ro_clip}). In this equation, \textit{f\textsubscript{c}} denotes the aforementioned threshold, while $q_{f_c}(\Bar{\mathbf{ro}}_{c})$ denotes the $f_c$\textsuperscript{th} quantile of the runoff vector in country \textit{c}. The values of this threshold (listed in Table \ref{tab:ror_thresholds}) are set such that the yearly average ROR capacity factors reaches 50\% across all countries\footnote{Considering a \textit{flood event threshold} of 0.9, the aggregate runoff time series within a region is clipped to its 90\textsuperscript{th} percentile. From a ROR plant design standpoint, this is equivalent to considering that a given plant is designed for a rated flow of \textit{p90} of the historical flow duration curve, an approach that is actually in use in the design of such units (though at different percentile values).}. Finally, the hourly availability of ROR plants can be obtained via (\ref{eq:ror_ro_pu}), where $K^{ROR}_c$ denotes the ROR installed capacity in country \textit{c}.

\begin{subequations}
\begin{align}
    ro_{c, t} &= \sum_{cell \in c} ro_{cell, t}, \hspace{2mm} \forall c \in \mathcal{C} \label{eq:ror_ro_sum}\\\vspace{2mm}
    \Bar{ro}_{c, t} &= \frac{ro_{c, t}}{max(\mathbf{ro}_{c})} , \hspace{2mm} \forall c \in \mathcal{C}, \hspace{1mm} \forall t \in \mathcal{T} \label{eq:ror_ro_norm}\\\vspace{2mm}
    \Tilde{ro}_{c,t} &= min(\Bar{ro}_{c,t}, \hspace{1mm} q_{f_c}(\Bar{\mathbf{ro}}_{c})), \hspace{2mm} \forall c \in \mathcal{C}, \hspace{1mm} \forall t \in \mathcal{T} \label{eq:ror_ro_clip}\\\vspace{2mm}
    p^{ROR}_{c,t} &= \Tilde{ro}_{c,t} \hspace{1mm} K^{ROR}_c, \hspace{2mm} \forall c \in \mathcal{C}, \hspace{1mm} \forall t \in \mathcal{T} \label{eq:ror_ro_pu}
\end{align}
\end{subequations}

\begin{table}
    \centering
	\renewcommand{\arraystretch}{1.1}
	\caption{Hydro run-of-river (ROR) flood event threshold values (p.u.) for various countries in Europe.}
	\label{tab:ror_thresholds}
    \begin{tabular}{cc||cc||cc}
    \toprule
    \textbf{ISO2} & \textbf{$f_c$} & \textbf{ISO2} & \textbf{$f_c$} & \textbf{ISO2} & \textbf{$f_c$} \\
    AT & 0.9  & GB & 0.85 & NO & 0.9 \\
    BE & 0.8  & GR & 0.9 & PL & 0.9 \\
    BG & 0.9  & HR & 0.9 & PT & 0.9 \\
    CH & 0.9  & HU & 0.9 & RO & 0.9 \\
    CZ & 0.9  & IE & 0.85 & RS & 0.9 \\
    DE & 0.7  & IT & 0.9 & SE & 0.9 \\
    ES & 0.9  & LT & 0.9 & SI & 0.9 \\
    FI & 0.9  & LU & 0.9 & SK & 0.9 \\
    FR & 0.9  & LV & 0.9 &  \\
    \bottomrule
    \end{tabular}
\end{table}

Reservoir-based hydro power plants (STO) are modelled as dispatchable units with limited generation capabilities, as their feed-in is usually limited by two aspects, i.e., water inflow availability and storage capabilities. At first, as STO reservoir capacity data is scarce in \cite{JRC_HY}, a procedure to approximate water retention capabilities on a country-by-country basis is employed. Subsequently, inflow estimation in the form of hourly time series is pursued. The accurate estimation of these two parameters prove fundamental in replicating the generation patterns of STO installations, which are often driven by seasonal fluctuations.

The assessment of country-based reservoir capacities develops in three steps. First, the results of a peer-reviewed hydropower modelling framework \cite{Hartel_2017} are queried. In case STO reservoir data is not available at this source for a specific country, information is sought for on the ENTSOE Transparency Platform \cite{ENTSOETransparency}, within time series of historical (i.e., 2014 to 2019) water reservoir levels. In this case, it is assumed that the maximum reported value for a country across the five available years coincides with its STO storage capabilities. Finally, if data is not found in any of the two aforementioned sources, the another database that stores more generic reservoir-specific data \cite{GrandDatabase} is queried. More explicitly, the STO storage capabilities of a given country are assumed equal to the sum of the capacities of all reservoirs in that country whose main usage is linked to hydroelectricity. Upon using the three aforementioned daata sources, \SI{189}{TWh} of storage capacity for reservoir-based plants are found within the European countries of interest.

The estimation of country-based inflows in STO reservoirs relies on reanalysis-based runoff time series \cite{ERA5}. However, instead of being expressed as per-unit values, the STO inflows are expressed in energy units (e.g., GWh). In this regard, let $\mathcal{G}$ denote the set of all reanalysis grid cells for which the inflow is computed. For each reanalysis grid cell $cell \in \mathcal{G}$, the product between the runoff and the corresponding grid cell surface area is computed, as seen in (\ref{eq:sto_flow_cell}). Given the fact that the runoff is natively expressed in meters\footnote{The runoff variable in use expresses "the depth the water would have if it were spread evenly over the grid box" \cite{ERA5_cds}.}, the product returns the equivalent volume of water available in $cell$ per unit time (i.e., water flow). Subsequently, the resulting time series are aggregated on a country basis, thus taking into account the grid cell geo-positioning  with respect to country borders (as per (\ref{eq:sto_flow_country})) and converted into an energy-based inflow via (\ref{eq:sto_flow_energy}). In the latter equation, $\rho_{H_2O}$ represents the water density (assumed \SI{1000}{km/m^3} and \textit{g} stands for the gravitational constant (i.e., \SI{9.81}{m/s^2}). Additionally, $h_c$ stands for the the country-average water head (assumed unitary in a first stage).

\begin{subequations}
\begin{align}
    q_{cell, t} &= ro_{cell, t} \hspace{1mm} \alpha_{cell}, \hspace{2mm} \forall cell \in \mathcal{G}, \forall t \in \mathcal{T} \label{eq:sto_flow_cell}\\\vspace{2mm}
    q_{c, t} &= \sum_{cell \in c} q_{cell, t}, \hspace{2mm} \forall c \in \mathcal{C}, \forall t \in \mathcal{T} \label{eq:sto_flow_country}\\\vspace{2mm}
    i^{STO, init}_{c,t} &= q_{c,t}, \hspace{1mm} \rho_{H_2O} \hspace{1mm} g \hspace{1mm} h_{c}, \hspace{2mm} \forall c \in \mathcal{C}, \forall t \in \mathcal{T} \label{eq:sto_flow_energy}\\\vspace{2mm}
    E^{STO}_c &= E^{HYDRO}_c - \sum_{t \in T} p^{ROR}_{c,t}, \hspace{2mm} \forall c \in \mathcal{C} \label{eq:sto_exp_prod}\\\vspace{2mm}
    fm_c &= \frac{E^{STO}_c}{\sum_{t \in T} i^{STO, init}_{c, t}}, \hspace{2mm} \forall c \in \mathcal{C} \label{eq:sto_flow_mult}\\\vspace{2mm}
    i^{STO}_{c,t} &= i^{STO, init}_{c,t} \hspace{1mm} fm_c, \hspace{2mm} \forall c \in \mathcal{C}, \forall t \in \mathcal{T} \label{eq:sto_inflow}
\end{align}
\end{subequations}

At this stage, the modelled STO water inflows are expressed in energy units, yet a brief comparison with historical yearly generation volumes reveals significant differences in favour of the latter. Indeed, on the one hand, a unitary water head is assumed in (\ref{eq:sto_flow_energy}), while the actual values of this parameter for individual plants vary between tens and hundreds of metres. On the other hand, it has been assumed that the entirety of the modelled inflows are used for hydroelectricity generation, whereas this is expected to be an unrealistic assumption, as e.g., part of the runoff is draining into the ground, part of it flows unconstrained through dam gates for environmental purposes, etc. In order to take these two factors (i.e., water head and the retain factor) into account in the definition of the STO inflow, a country-specific \textit{flow multiplier} or $fm_c$ is derived. For each country, yearly integrated ROR production data (whose estimation has been previously discussed) is subtracted from the total hydroelectricity generation volumes ($E_c^{HYDRO}$ obtained from \cite{eurostat_hydro}) and the remainder is attributed to reservoir-based plants, as described by (\ref{eq:sto_exp_prod}). The ratio (\ref{eq:sto_flow_mult}) between the expected generation and the yearly integrated STO inflows assuming a unitary water head defines the flow multiplier mentioned above and centralized in Table \ref{tab:flow_multiplier}. The corrected STO inflow time series are therefore obtained by multiplying their initial values with this country-specific scalar, as seen in Eq. (\ref{eq:sto_inflow}).

\begin{table}
    \centering
	\renewcommand{\arraystretch}{1.1}
	\caption{Reservoir-based hydro (STO) flow multiplier values (unitless) for various countries in Europe.}
	\label{tab:flow_multiplier}
    \begin{tabular}{cc||cc||cc}
    \toprule
    \textbf{ISO2} & \textbf{\textit{$fm_c$}} & \textbf{ISO2} & \textbf{\textit{$fm_c$}} & \textbf{ISO2} & \textbf{\textit{$fm_c$}} \\
    AT & 114.3  & GB & 9.8 & NO & 279.3 \\
    BE & 64.3  & GR & 82.7 & PL & 33.9 \\
    BG & 116.9  & HR & 123.9 & PT & 194.5 \\
    CH & 191.3  & HU & 11.3 & RO & 168.1 \\
    CZ & 96.3  & IE & 0.07 & RS & 392.6 \\
    DE & 48.7  & IT & 83.9 & SE & 159.9 \\
    ES & 279.2  & LT & 44.4 & SI & 103.1 \\
    FI & 22.7 & LU & 499.1 & SK & 129.5 \\
    FR & 118.9  & LV & 3.6 &  \\
    \bottomrule
    \end{tabular}
\end{table}

In the exercise at hand, pumped-hydro storage (PHS) is available as storage technology without the possibility of expansion. PHS units are modelled without natural inflows (i.e., closed-loop schemes) and with a unitary self-discharge efficiency. Additionally, it is assumed that, for all PHS units, both turbine and pumping modes have identical rated powers and efficiencies. Installed capacities of such units are retrieved from \cite{JRC_HY} yet, as for STO units, information regarding to their storage capabilities is scarce. However, an accurate assessment of such information is paramount for an accurate modelling of PHS plants which, with over \SI{54}{GW} of installed capacity, have a significant impact in the integration of RES throughout Europe. In this regard, the following unit-based approach is proposed. At first, storage capabilities data is sought for each individual plant obtained from \cite{JRC_HY} within a peer-reviewed database \cite{Geth2015}. If for a given plant, unit-specific data does not exist, but country-specific duration (i.e., energy-to-power ratio) data is available, the latter is used to approximate storage capacities starting from the installed capacities. In case country- and plant-specific data is missing, a default duration of \SI{6}{h} is used to derive pumped-hydro storage capabilities. This procedure results in a total European pumped-hydro storage potential of \SI{1930}{GWh} whose distribution is presented in Table \ref{tab:phs_stor}.

\begin{table}
    \centering
	\renewcommand{\arraystretch}{1.1}
	\caption{Pumped-hydro storage (PHS) power and energy capacities for various countries in Europe.}
	\label{tab:phs_stor}
    \begin{tabular}{ccc||ccc||ccc}
    \toprule
    \textbf{ISO2} & \textbf{\textit{$p_c$}} & \textbf{\textit{$e_c$}} & \textbf{ISO2} & \textbf{\textit{$p_c$}} & \textbf{\textit{$e_c$}} & \textbf{ISO2} & \textbf{\textit{$p_c$}} & \textbf{\textit{$e_c$}} \\
    & [GW] & [GWh] && [GW] & [GWh] && [GW] & [GWh] \\
    AT & 3.6 & 159.4 & GB & 2.9 & 26.7 & NO & 1.3 & 472.6 \\
    BE & 1.3 & 5.7 & GR & 0.7 & 5.1 & PL & 1.7 & 7.3 \\
    BG & 1.4 & 41.1 & HR & 0.5 & 4.4 & PT & 3.9 & 130.7 \\
    CH & 4.5 & 648.5 & HU &&& RO &&\\
    CZ & 1.2 & 5.57 & IE & 0.3 & 1.8 & RS & 0.6 & 3.6 \\
    DE & 7.8 & 45.6 & IT & 7.9 & 81.0 & SE & 0.1 & 116.5\\
    ES & 7.9 & 83.1 & LT & 0.9 & 10.8 & SI & 0.2 & 0.5 \\
    FI &&& LU &&& SK & 1.0 & 4.6 \\
    FR & 5.3 & 85.6 & LV &&&  \\
    \bottomrule
    \end{tabular}
\end{table}

\subsection*{Techno-economic assumptions}\label{Assumptions}
\subsubsection*{Economic parameters}
Economic parameters of every generation, storage or transmission technology are displayed in Table \ref{tab:tech-costs}. USD/EUR conversions are made assuming a conversion rate of 0.8929 EUR per USD. Then, GBP/EUR conversions are made assuming a conversion rate of 1.1405 EUR per GBP. Similarly, a AUD/EUR conversion rate of 0.6209 EUR per AUD is considered.

\begin{table}
\centering
\renewcommand{\arraystretch}{1.2}
\begin{threeparttable}
\begin{tabular}{lccccc}
\toprule
Plant              & CAPEX                & FOM        & VOM      & Lifetime           & Source  \\
                  & M\euro/GW(h)          & M\euro$\mathrm{/GW \times yr}$ & M\euro/GWh & years &         \\\midrule
Onshore wind       & N/A              & 29.47      & 0.00     & 25                 &  \cite{ATB2020}       \\
Offshore wind      & 1881.08              & 49.11      & 0.00     & 25                 &   \cite{ATB2020}      \\
Utility-scale PV   & N/A               & 7.14       & 0.00     & 25                 &    \cite{ATB2020}     \\
Distributed PV     & N/A               & 5.36       & 0.00     & 25                 &    \cite{ATB2020}     \\
OCGT               & 838.87               & 3.03       & 0.0076   & 30                 &    \cite{AEMO2020}     \\
CCGT               & 1005.27              & 7.58       & 0.0053   & 30                 &    \cite{AEMO2020}     \\
Nuclear            & N/A                  & 106.25     & 0.0018   & 40                 &    \cite{ATB2020}     \\
Run-of-river hydro & N/A                  & 0.00       & 0.0119   & N/A                &    \cite{IEA2020Costs}     \\
Reservoir hydro    & N/A                  & 0.00       & 0.0152   & N/A                &    \cite{IEA2020Costs}     \\
Li-Ion (power)     & 100                  & 0.54       & N/A      & 10\tnote{1} &  \cite{DEA2020STO}  \\
Li-Ion (energy)    & 94   & N/A        & 0.0017   & 10\tnote{1} &  \cite{DEA2020STO}  \\
Pumped-hydro storage      & N/A                  & 14.20      & 0.0002   & N/A                &  \cite{HWires}      \\
HVAC               & 2.22\tnote{2} & 0.017      & N/A      & 40\tnote{1} &  \cite{RealiseGrid} \\
HVDC               & 1.76\tnote{3} & 0.021      & N/A      & 40\tnote{1} &  \cite{RealiseGrid} \\
\bottomrule
\end{tabular}
    \begin{tablenotes}\footnotesize
    \item[1] Assumed value.
    \item[2] Expressed per km and derived from a single circuit 1000 MVA, 400 kV cable.
    \item[3] Expressed per km and derived from a 1100 MW, 350 kV underground DC cable pair.
    \end{tablenotes}
    \caption{Economic parameters of generation, storage and transmission technologies.}
    \label{tab:tech-costs}
\end{threeparttable}

\end{table}

\subsubsection*{Technical parameters}
For generation technologies, efficiencies represent the ratio between primary energy input and electricity output. For storage technologies, three efficiencies are provided, i.e. for the discharging (D) and charging (S) states, as well as the self-discharge (SD) efficiency (i.e., one minus this value corresponds to the state-of-charge internal losses). For transmission technologies, the efficiency is provided per 1000 km. Besides efficiency values, some technologies are modelled with hourly ramp rates (up- and down-regulation), as well as with minimum must-run levels. At this stage the only technology with such constraints is the nuclear power generator, i.e., those generators can ramp their production up or down by a maximum of 10\% per hour and must always output power at a minimum of 80\% of their capacity. All values are centralized in Table \ref{tab:tech-eff}.

\begin{table}[ht]
    \centering
    \renewcommand{\arraystretch}{1.2}
    \begin{threeparttable}
    \begin{tabular}{llcccccc}
        \toprule
        Plant & \multicolumn{3}{c}{Efficiency} & \multicolumn{2}{c}{Ramp rates\tnote{1}} & Must-run\tnote{1}\\
        & D & S & SD & Up & Down &\\
        & [\%] & [\%] & [\%] & [\%/hr] & [\%/hr] & [\%]\\
        \midrule
        OCGT & 41.0\tnote{2} & & & & \\
        CCGT & 58.0\tnote{2} & & & & \\
        Nuclear & 36.0\tnote{3} & & & 10.0 & 10.0 & 80.0\\
        Run-of-river hydro & 85.0\tnote{1} & & & &\\
        Reservoir hydro & 85.0\tnote{1} & & & &\\
        Pumped-hydro storage & 90.0\tnote{4} & 90.0\tnote{4} & 100.0\tnote{1} & &\\
        Li-Ion storage & 93.0\tnote{5} & 93.0\tnote{5} & 99.5\tnote{1} & &\\
        HVAC & 93.0\tnote{6} & & & &\\
        HVDC & 97.0\tnote{6} & & & &\\
        \bottomrule
    \end{tabular}
    \begin{tablenotes}\footnotesize
    \item[1] Assumed values.
    \item[2] Retrieved from \cite{AEMO2020}.
    \item[3] Retrieved from \cite{ATB2020}.
    \item[4] Retrieved from \cite{SANDIA2013}.
    \item[5] Retrieved from \cite{HWires}.
    \item[6] Retrieved from \cite{ETSAP_TRANSMISSION}.
    \end{tablenotes}
    \end{threeparttable}
    \caption{Generation, storage and transmission technologies operational parameters.}
    \label{tab:tech-eff}
\end{table}

Some technologies (OCGT and CCGT) are using fuels which release CO\textsubscript{2} when burnt. Specific CO\textsubscript{2} emissions and associated feedstock costs of different fuels are collected in Table \ref{tab:tech-fuels}.

\begin{table}[ht]
    \centering
    \begin{threeparttable}
    \begin{tabular}{llcc}
        \toprule
        Plant & Fuel & Fuel Cost & CO\textsubscript{2}\\
        && \euro$\mathrm{/MWh_{th}}$ & $\mathrm{ton/MWh_{th}}$ \\
        \midrule
        OCGT & gas & 26.5\tnote{1} & 0.225\tnote{2}\\
        CCGT & gas & 26.5\tnote{1} & 0.225\tnote{2}\\
        Nuclear & uranium & 1.7\tnote{1} & 0.0\\
        \bottomrule
    \end{tabular}
        \begin{tablenotes}\footnotesize
    \item[1] Values are obtained from the TYNDP2020 Scenario Report, reference year 2040 \cite{TYNDP2020}.
    \item[2] Values for stationary combustion from the IPCC Emissions Factor Database \cite{IPCC2006}.
    \end{tablenotes}
    \end{threeparttable}
    \caption{Fuels and associated costs and specific emissions.}
    \label{tab:tech-fuels}
\end{table}

The capacity expansion framework includes a carbon budget constraint reflecting the total amount of CO\textsubscript{2} that can be emitted by the underlying power system over one year. In line with the latest climate agreements, this budget is enforced as a share of the EU-wide 1990 emission levels. Data collection for yearly emission levels relies on carbon intensity information released by the Europe Environment Agency (EEA) \cite{EEA2020} and on yearly electricity generation data from the International Energy Agency (IEA) \cite{IEA2020}. The baseline cost of CO\textsubscript{2} is set to 41.85 \euro$\mathrm{/tCO_2}$ and it reflects its price on the EU ETS on March 30, 2021.

\clearpage
\section*{Nomenclature}

\subsection*{Indices \& Sets}
\begin{description}[leftmargin=4em,style=nextline]
    \item[$c, \mathcal{C}$] line, set of transmission corridors, $\mathcal{C} \subset \mathcal{N}_B \times \mathcal{N}_B$
    \item[$\mathcal{C}_n^{+}$, $\mathcal{C}_n^{-}$] set of inbound links into node $n$, with $\mathcal{C}_n^{+} = \{c \in \mathcal{C} | c = (u, n), u \in \mathcal{N}_n^{+} \}$, where $\mathcal{N}_n^{+} = \{u \in \mathcal{N}_B | (u, n) \in \mathcal{C}\}$, set of outbound links from node $n$, with $\mathcal{C}_n^{-} = \{c \in \mathcal{C} | c = (n, v), v \in \mathcal{N}_n^{-} \}$, where $\mathcal{N}_n^{-} = \{v \in \mathcal{N}_B | (n, v) \in \mathcal{C}\}$
    \item[$\mathcal{D}$] set of technologies providing firm (i.e., non-variable) capacity
    \item[$g, \mathcal{G}$] conventional generation technology and the associated set
        \item[$l, L_n$] RES site, set of sited RES locations associated with topology node $n$
    \item[$\mathcal{L}_{0}$] set of legacy RES sites
    \item[$\mathcal{L}, \mathcal{L}_n$] set of candidate sites, set of candidate sites in partition $n \in \mathcal{N}_B$
    \item[$n, \mathcal{N}_B$] node, set of topology nodes
    \item[$r, \mathcal{R}$] RES technology and set of RES technologies
    \item[$s, \mathcal{S}$] storage technology and set of storage technologies
    \item[$t, \mathcal{T}$] time index, set of time periods
    \item[$w, \mathcal{W}$] time window, set of time windows
\end{description}

\subsection*{Parameters}
\begin{description}[leftmargin=4.5em,style=nextline]
        \item[$A_{nl}$] incidence matrix entry indicating whether location $l$ belongs to partition $\mathcal{L}_n$
        \item[$\alpha_{l\mathrm{w}}$] reference production level at location $l$ during time window $w$
        \item[$c$] threshold defining the global coverage of time windows
        \item[$D_{lw}$] criticality matrix entry indicating whether location $l$ covers time window $w$
        \item[$\delta$] time window length
		\item[$\Delta_g^{+}, \Delta_g^{i}$] upward and downward ramp-rate of conventional generation technology $g$
		\item[$\epsilon_{site}$] reference site surface utilization factor
		\item[$\eta_g$] thermal efficiency of generation technology $g \in \mathcal{G}$
		\item[$\eta^{SD/D/C}_s$] self-discharging, discharging and charging efficiency of storage technology $s \in \mathcal{S}$
		\item[$k_n$] number of desired deployments in partition $\mathcal{L}_n$
		\item[$\underline{\kappa}_{c}$] initial capacity for transmission line $c \in \mathcal{C}$
		\item[$\underline{\kappa}_{l}$] initial capacity at site $l \in \{\mathcal{L}\}$
		\item[$\underline{\kappa}_{nx}$] initial capacity for technology $x \in \{\mathcal{G}, \mathcal{R}, \mathcal{S}\} at node \textit{n}$
		\item[$\Bar{\kappa}_{c}$] maximum allowable installed capacity of line $c \in \mathcal{C}$
	    \item[$\Bar{\kappa}_{l}$] maximum allowable installed capacity at site $l \in \{\mathcal{L}\}$
	    \item[$\Bar{\kappa}_{nx}$] maximum allowable installed capacity of technology $x \in \{\mathcal{G}, \mathcal{R}, \mathcal{S}\} at node \textit{n}$
	    \item[$\kappa_n$] target capacity at partition $\mathcal{L}_n$
	    \item[$g_l$] resource assessment measure at location $l$
	    \item[$h_l$] RES transfer function at location $l$
	    \item[$I$] number of iterations in the simulated annealing algorithm
	    \item[$\lambda_{nt}$] electricity demand at node \textit{n} and time \textit{t}
	    \item[$\Bar{\lambda}_{nt}$] electricity demand during time window $w$ at node $n$
	    \item[$\hat{\lambda}_{n}$] peak electricity demand at node \textit{n}
		\item[$\mu_j$] minimum required operational level of technology $j \in \{\mathcal{G}, \mathcal{S}\}$
		\item[$N$] number of epochs in the simulated annealing algorithm
		\item[$\nu_g^{CO_2}$] specific CO\textsubscript{2} emissions associated with generation technology $g \in \mathcal{G}$
	    \item[$\pi_{lt}$] per-unit availability (i.e., capacity factor) of RES site \textit{l} and time \textit{t}
	    \item[$\Bar{\pi}_{l\mathrm{w}}$] per-unit availability (i.e., capacity factor) of RES site $l$ during time window $w$
	    \item[$\tilde{\pi}_{l}$] average per-unit availability (i.e., capacity factor) of RES site $l$
	    \item[$\Pi_{l}$] capacity credit of RES site $l \in \mathcal{L}$
	    \item[$\Pi_{nr}$] capacity credit of RES technology $r$ at node $n$
	    \item[$\phi_s$] charge-to-discharge ratio of storage technology $s \in \mathcal{S}$
	    \item[$\Phi_n$] planning reserve margin at node $n \in \mathcal{N}_B$
		\item[$q^{CO_2}_{ngt}$] CO\textsubscript{2} emissions resulting from the operation of generation technology \textit{g} at node $n$ and time $t$
		\item[$r$] neighbourhood radius in the simulated annlealing algorithm
		\item[$\rho_r$] power density of RES technology $r \in \mathcal{R}$
		\item[$s_{lt}$] renewable resource (e.g., wind speed, solar irradiation) at location $l$ and time $t$
		\item[$T(i)$] annealing schedule (function of iteration $i$)
        \item[$\theta^{ens}$] economic penalty for demand curtailment
        \item[$\theta^l_f, \theta^l_v$] fixed (FOM) and variable (VOM) operation and maintenance cost of RES site $l \in \{\mathcal{L}\}$
		\item[$\theta^j_f, \theta^j_v$] fixed (FOM) and variable (VOM) operation and maintenance cost of technology $j \in \{\mathcal{G}, \mathcal{R}, \mathcal{S}, \mathcal{C}\}$
		\item[$\Psi^{CO_2}$] system-wide CO\textsubscript{2} budget
		\item[$\sigma_{site}$] reference site surface area
		\item[$\varsigma$] dimensionless scalar
		\item[$\omega_t$] weight of each operating condition $t \in \mathcal{T}$ in the objective function and CO\textsubscript{2} emissions
		\item[$\omega_s$] weight of each operating condition $t \in \mathcal{T}$ in the operation of storage units
		\item[$\zeta^l$] annualized investment cost at RES site $l \in \{\mathcal{L}\}$
		\item[$\zeta^j$] annualized investment cost of technology $j \in \{\mathcal{G}, \mathcal{R}, \mathcal{S}, \mathcal{C}\}$
		\item[$\zeta^s_S$] annualized investment cost of the energy component of storage technology $s \in \{\mathcal{S}\}$
\end{description}

\subsection*{Variables}
\begin{description}[leftmargin=5.5em,style=nextline]
    \item[$K_{c} \in \mathbb{R}_+$] installed capacity of transmission line $c \in \mathcal{C}$
    \item[$K_{l} \in \mathbb{R}_+$] installed capacity of RES site $l \in \mathcal{L}$
    \item[$K_{nj} \in \mathbb{R}_+$] installed capacity of generation tech. $j \in \{\mathcal{G}, \mathcal{R}\}$ at node \textit{n}
    \item[$K_{ns} \in \mathbb{R}_+$] power component rated capacity of storage technology $s \in \mathcal{S}$
    \item[$p^{C/D}_{nst} \in \mathbb{R}_+$] charging/discharging flow of storage technology $s \in \mathcal{S}$, at node \textit{n} and time \textit{t}
    \item[$p_{ct} \in \mathbb{R}$] power flow over line $l \in \mathcal{C}$ at time \textit{t}
    \item[$p^{ens}_{nt} \in \mathbb{R}_+$] unserved demand at node $n \in \mathcal{N}_B$ and time \textit{t}
    \item[$p_{ngt} \in \mathbb{R}_+$] in-feed of generation technology $g \in \mathcal{G}$, at node \textit{n} and time \textit{t}
    \item[$p_{lt} \in \mathbb{R}_+$] in-feed of RES site $l \in L_n$ at time \textit{t}
    \item[$S_{ns} \in \mathbb{R}_+$] energy component rated capacity of storage technology $s \in \mathcal{S}$
    \item[$x_{l} \in \{0,1\}$] binary variable indicating deployment of site $l$
    \item[$y_{w} \in \{0,1\}$] binary variable indicating the system-wide criticality of window $w$
\end{description}

\bibliographystyle{elsarticle-num}
\bibliography{main}

\begin{thebibliography}{10}
\expandafter\ifx\csname url\endcsname\relax
  \def\url#1{\texttt{#1}}\fi
\expandafter\ifx\csname urlprefix\endcsname\relax\def\urlprefix{URL }\fi
\expandafter\ifx\csname href\endcsname\relax
  \def\href#1#2{#2} \def\path#1{#1}\fi

\bibitem{IEA2020b}
{International Energy Agency (IEA)},
  \href{https://www.iea.org/articles/renewables-2020-data-explorer}{{Renewables
  2020 Data Explorer}} (2020).
\newline\urlprefix\url{https://www.iea.org/articles/renewables-2020-data-explorer}

\bibitem{IEA2050}
\href{https://www.iea.org/reports/net-zero-by-2050}{{Net Zero by 2050 - A
  Roadmap for the Global Energy Sector}}, Tech. rep., International Energy
  Agency (2021).
\newline\urlprefix\url{https://www.iea.org/reports/net-zero-by-2050}

\bibitem{Segreto2020}
M.~Segreto, L.~Principe, A.~Desormeaux, M.~Torre, L.~Tomassetti, P.~Tratzi,
  V.~Paolini, F.~Petracchini, {Trends in Social Acceptance of Renewable Energy
  Across Europe—A Literature Review}, International Journal of Environmental
  Research and Public Health 17~(24) (2020).
\newblock \href {https://doi.org/10.3390/ijerph17249161}
  {\path{doi:10.3390/ijerph17249161}}.

\bibitem{IRENA2020}
\href{https://www.irena.org/publications/2020/Jun/Renewable-Power-Costs-in-2019}{{Renewable
  Power Generation Costs in 2019}}, Tech. rep., International Renewable Energy
  Agency (2020).
\newline\urlprefix\url{https://www.irena.org/publications/2020/Jun/Renewable-Power-Costs-in-2019}

\bibitem{globalwindatlas}
J.~Badger, I.~Bauwens, P.~Casso, N.~Davis, A.~Hahmann, S.~Bo~Krohn~Hansen,
  B.~Ohrbeck~Hansen, D.~Heathfield, J.~O. Knight, O.~Lacave, G.~Lizcano,
  A.~Bosch~i Mas, N.~Gylling~Mortensen, B.~T. Olsen, M.~Onninen, A.~Potter~van
  Loon, P.~Volker, \href{https://globalwindatlas.info/}{{Global Wind Atlas
  3.0}} (2021).
\newline\urlprefix\url{https://globalwindatlas.info/}

\bibitem{EC_2018}
\href{https://ec.europa.eu/clima/policies/strategies/2050_en}{{A Clean Planet
  for all - A European strategic long-term vision for a prosperous, modern,
  competitive and climate neutral economy}}, Tech. rep., European Commission
  (2018).
\newline\urlprefix\url{https://ec.europa.eu/clima/policies/strategies/2050_en}

\bibitem{WindEurope2020_main}
K.~Freeman, C.~Frost, G.~Hundleby, A.~Roberts, B.~Valpy, H.~Holttinen,
  L.~Ramirez, I.~Pineda,
  \href{https://windeurope.org/wp-content/uploads/files/about-wind/reports/WindEurope-Our-Energy-Our-Future.pdf}{{Our
  energy, our future - How offshore wind will help {E}urope to go
  carbon-neutral}}, Tech. rep., WindEurope (2019).
\newline\urlprefix\url{https://windeurope.org/wp-content/uploads/files/about-wind/reports/WindEurope-Our-Energy-Our-Future.pdf}

\bibitem{Engeland2017}
K.~Engeland, M.~Borga, J.-D. Creutin, B.~François, M.-H. Ramos, J.-P. Vidal,
  {Space-Time Variability of Climate Variables and Intermittent Renewable
  Electricity Production - A Review}, Renewable and Sustainable Energy Reviews
  79 (2017) 600 -- 617.
\newblock \href {https://doi.org/https://doi.org/10.1016/j.rser.2017.05.046}
  {\path{doi:https://doi.org/10.1016/j.rser.2017.05.046}}.

\bibitem{Geth2015}
F.~Geth, T.~Brijs, J.~Kathan, J.~Driesen, R.~Belmans, {An Overview of
  Large-Scale Stationary Electricity Storage Plants in Europe: Current Status
  and New Developments}, Renewable and Sustainable Energy Reviews 52 (2015)
  1212--1227.
\newblock \href {https://doi.org/https://doi.org/10.1016/j.rser.2015.07.145}
  {\path{doi:https://doi.org/10.1016/j.rser.2015.07.145}}.

\bibitem{Stenclik2017}
D.~Stenclik, P.~Denholm, B.~Chalamala, {Maintaining Balance: The Increasing
  Role of Energy Storage for Renewable Integration}, IEEE Power and Energy
  Magazine 15~(6) (2017) 31--39.
\newblock \href {https://doi.org/https://doi.org/10.1109/MPE.2017.2729098}
  {\path{doi:https://doi.org/10.1109/MPE.2017.2729098}}.

\bibitem{OConnell2014}
N.~O'Connell, P.~Pinson, H.~Madsen, M.~O'Malley, {Benefits and Challenges of
  Electrical Demand Response: A Critical Review}, Renewable and Sustainable
  Energy Reviews 39 (2014) 686--699.
\newblock \href {https://doi.org/https://doi.org/10.1016/j.rser.2014.07.098}
  {\path{doi:https://doi.org/10.1016/j.rser.2014.07.098}}.

\bibitem{Milligan1999}
M.~R. Milligan, R.~Artig, \href{https://www.osti.gov/biblio/750939}{{Choosing
  Wind Power Plant Locations and Sizes Based on Electric Reliability Measures
  Using Multiple-Year Wind Speed Measurements}}, Tech. Rep. CP-500-26724, NREL
  (1999).
\newline\urlprefix\url{https://www.osti.gov/biblio/750939}

\bibitem{Giebel2001}
G.~Giebel, {On the Benefits of Distributed Generation of Wind Energy in
  Europe}, {PhD Thesis}, {University of Oldenburg, Germany} (Jul 2001).

\bibitem{Jurasz2020}
J.~Jurasz, F.~Canales, A.~Kies, M.~Guezgouz, A.~Beluco, {A Review on the
  Complementarity of Renewable Energy Sources: Concept, Metrics, Application
  and Future Research Directions}, Solar Energy 195 (2020) 703--724.
\newblock \href {https://doi.org/https://doi.org/10.1016/j.solener.2019.11.087}
  {\path{doi:https://doi.org/10.1016/j.solener.2019.11.087}}.

\bibitem{MacDonald2016}
A.~MacDonald, C.~Clack, A.~Alexander, A.~Dunbar, J.~Wilczak, Y.~Xie, Future
  cost-competitive electricity systems and their impact on {US $CO_2$}
  emissions, Nature Climate Change 6 (2016).
\newblock \href {https://doi.org/doi:10.1038/NCLIMATE2921}
  {\path{doi:doi:10.1038/NCLIMATE2921}}.

\bibitem{Berger2018}
M.~Berger, D.Radu, R.~Fonteneau, R.~Henry, M.~Glavic, X.~Fettweis, M.~L. Du,
  P.~Panciatici, L.~Balea, D.~Ernst,
  \href{https://www.sciencedirect.com/science/article/abs/pii/S0360544220304151}{Critical
  time windows for ren. resource complementarity assessment}, Energy 198
  (2020).
\newblock \href {https://doi.org/10.1016/j.energy.2020.117308}
  {\path{doi:10.1016/j.energy.2020.117308}}.
\newline\urlprefix\url{https://www.sciencedirect.com/science/article/abs/pii/S0360544220304151}

\bibitem{ERA5}
{European {C}entre for {M}edium-{R}ange {W}eather {F}orecasts - ECMWF},
  \href{https://confluence.ecmwf.int//display/CKB/}{Copernicus knowledge base -
  {ERA}5 data documentation} (2018).
\newline\urlprefix\url{https://confluence.ecmwf.int//display/CKB/}

\bibitem{Pfenninger2014}
S.~Pfenninger, A.~Hawkes, J.~Keirstead,
  \href{http://www.sciencedirect.com/science/article/pii/S1364032114000872}{Energy
  systems modeling for twenty-first century energy challenges}, Renewable and
  Sustainable Energy Reviews 33 (2014) 74 -- 86.
\newblock \href {https://doi.org/https://doi.org/10.1016/j.rser.2014.02.003}
  {\path{doi:https://doi.org/10.1016/j.rser.2014.02.003}}.
\newline\urlprefix\url{http://www.sciencedirect.com/science/article/pii/S1364032114000872}

\bibitem{Krishnan2016}
V.~{Krishnan}, W.~{Cole}, Evaluating the value of high spatial resolution in
  national {Capacity Expansion Models using ReEDS}, {2016 IEEE PES General
  Meeting} (2016).

\bibitem{Brown2021}
M.~M. Frysztacki, J.~Hörsch, V.~Hagenmeyer, T.~Brown,
  \href{https://www.sciencedirect.com/science/article/pii/S0306261921002439}{The
  strong effect of network resolution on electricity system models with high
  shares of wind and solar}, Applied Energy 291 (2021) 116726.
\newblock \href
  {https://doi.org/https://doi.org/10.1016/j.apenergy.2021.116726}
  {\path{doi:https://doi.org/10.1016/j.apenergy.2021.116726}}.
\newline\urlprefix\url{https://www.sciencedirect.com/science/article/pii/S0306261921002439}

\bibitem{Jerez2015}
S.~Jerez, F.~Thais, I.~Tobin, M.~Wild, A.~Colette, P.~Yiou, R.~Vautard, The
  {CLIMIX} model: A tool to create and evaluate spatially-resolved scenarios of
  photovoltaic and wind power development, Renewable and Sustainable Energy
  Reviews 42 (2015) 1--15.
\newblock \href {https://doi.org/doi:10.1016/j.rser.2014.09.041}
  {\path{doi:doi:10.1016/j.rser.2014.09.041}}.

\bibitem{Becker2018}
R.~Becker, D.~Thrän, \href{http://dx.doi.org/10.3390/en11040978}{Optimal
  siting of wind farms in wind energy dominated power systems}, Energies 11~(4)
  (2018) 978.
\newblock \href {https://doi.org/10.3390/en11040978}
  {\path{doi:10.3390/en11040978}}.
\newline\urlprefix\url{http://dx.doi.org/10.3390/en11040978}

\bibitem{Musselman2019}
A.~Musselman, V.~M. Thomas, N.~Boland, D.~Nazzal, Optimizing wind farm siting
  to reduce power system impacts of wind variability, Wind Energy 22~(7) (2019)
  894--907.
\newblock \href {https://doi.org/https://doi.org/10.1002/we.2328}
  {\path{doi:https://doi.org/10.1002/we.2328}}.

\bibitem{Hu2019}
J.~Hu, R.~Harmsen, W.~Crijns-Graus, E.~Worrell, Geographical optimization of
  variable renewable energy capacity in {C}hina using modern portfolio theory,
  Applied Energy 253 (2019) 113614.
\newblock \href
  {https://doi.org/https://doi.org/10.1016/j.apenergy.2019.113614}
  {\path{doi:https://doi.org/10.1016/j.apenergy.2019.113614}}.

\bibitem{Berger2020}
M.~{Berger}, D.~{Radu}, A.~{Dubois}, Y.~{Dvorkin}, H.~{Pandzic}, Q.~{Louveaux},
  D.~{Ernst}, \href{https://orbi.uliege.be/handle/2268/251037}{Siting renewable
  power generation assets with combinatorial optimisation}, pre-print (2020).
\newline\urlprefix\url{https://orbi.uliege.be/handle/2268/251037}

\bibitem{Dagoumas2018}
N.~E. Koltsaklis, A.~S. Dagoumas, State-of-the-art generation expansion
  planning: A review, Applied Energy 230 (2018).
\newblock \href {https://doi.org/doi:10.1016/j.apenergy.2018.08.087}
  {\path{doi:doi:10.1016/j.apenergy.2018.08.087}}.

\bibitem{Baringo2014}
L.~Baringo, A.~J. Conejo, Strategic wind power investment, IEEE Transactions on
  Power Systems 29~(3) (2014) 1250--1260.
\newblock \href {https://doi.org/10.1109/TPWRS.2013.2292859}
  {\path{doi:10.1109/TPWRS.2013.2292859}}.

\bibitem{Munoz2014}
F.~D. {Munoz}, B.~F. {Hobbs}, J.~L. {Ho}, S.~{Kasina}, An engineering-economic
  approach to transmission planning under market and regulatory uncertainties:
  {WECC} case study, IEEE Transactions on Power Systems 29~(1) (2014) 307--317.
\newblock \href {https://doi.org/10.1109/TPWRS.2013.2279654}
  {\path{doi:10.1109/TPWRS.2013.2279654}}.

\bibitem{Zappa2019}
W.~Zappa, M.~Junginger, M.~van~den Broek, Is a 100\% renewable {E}uropean power
  system feasible by 2050?, Applied Energy 233-234 (2019) 1027--1050.
\newblock \href {https://doi.org/doi:10.1016/j.apenergy.2018.08.109}
  {\path{doi:doi:10.1016/j.apenergy.2018.08.109}}.

\bibitem{Kotzur2018}
L.~Kotzur, P.~Markewitz, M.~Robinius, D.~Stolten, Impact of different time
  series aggregation methods on optimal energy system design, Renewable Energy
  117 (2018) 474 -- 487.
\newblock \href {https://doi.org/https://doi.org/10.1016/j.renene.2017.10.017}
  {\path{doi:https://doi.org/10.1016/j.renene.2017.10.017}}.

\bibitem{Callaway2017}
G.~Wu, R.~Deshmukh, K.~Ndhlukulac, T.~Radojicic, J.~Reilly-Moman, A.~Phadke,
  D.~Kammen, D.~Callaway, Strategic siting and regional grid interconnections
  key to low-carbon futures in {A}frican countries, Proceedings of the National
  Academy of Sciences 114 (2017).
\newblock \href {https://doi.org/doi:10.1073/pnas.1611845114}
  {\path{doi:doi:10.1073/pnas.1611845114}}.

\bibitem{Zappa2018}
W.~Zappa, M.~van~den Broek, Analysing the potential of integrating wind and
  solar power in {E}urope using spatial optimisation under various scenarios,
  Renewable and Sustainable Energy Reviews 94 (2018) 1192 -- 1216.
\newblock \href {https://doi.org/doi:10.1016/j.rser.2018.05.071}
  {\path{doi:doi:10.1016/j.rser.2018.05.071}}.

\bibitem{Bertsimas1993}
D.~Bertsimas, J.~Tsitsiklis, Simulated annealing, Statist. Sci. 8~(1) (1993)
  10--15.
\newblock \href {https://doi.org/10.1214/ss/1177011077}
  {\path{doi:10.1214/ss/1177011077}}.

\bibitem{Mertens2018}
T.~Mertens, K.~Bruninx, J.~Duerinck, E.~Delarue,
  \href{https://www.mech.kuleuven.be/en/tme/research/energy_environment/Pdf/wp-en2018-16}{The
  impact of planning reserve margins and demand uncertainty in generation
  expansion models}, IAEE Proceedings (2018).
\newline\urlprefix\url{https://www.mech.kuleuven.be/en/tme/research/energy_environment/Pdf/wp-en2018-16}

\bibitem{Milligan2008}
C.~Ensslin, M.~Milligan, H.~Holttinen, M.~O'Malley, A.~Keane, Current methods
  to calculate capacity credit of wind power, {IEA} collaboration, 2008, pp. 1
  -- 3.
\newblock \href {https://doi.org/10.1109/PES.2008.4596006}
  {\path{doi:10.1109/PES.2008.4596006}}.

\bibitem{PyPSA}
T.~Brown, J.~H\"orsch, D.~Schlachtberger,
  \href{https://doi.org/10.5334/jors.188}{{PyPSA: Python for Power System
  Analysis}}, Journal of Open Research Software 6 (2018).
\newblock \href {http://arxiv.org/abs/1707.09913} {\path{arXiv:1707.09913}},
  \href {https://doi.org/10.5334/jors.188} {\path{doi:10.5334/jors.188}}.
\newline\urlprefix\url{https://doi.org/10.5334/jors.188}

\bibitem{Staffell2016Solar}
I.~Staffell, S.~Pfenninger,
  \href{http://www.sciencedirect.com/science/article/pii/S0360544216311811}{Using
  bias-corrected reanalysis to simulate current and future wind power output},
  Energy 114 (2016) 1224 -- 1239.
\newblock \href {https://doi.org/https://doi.org/10.1016/j.energy.2016.08.068}
  {\path{doi:https://doi.org/10.1016/j.energy.2016.08.068}}.
\newline\urlprefix\url{http://www.sciencedirect.com/science/article/pii/S0360544216311811}

\bibitem{Pffenninger2016Wind}
S.~Pfenninger, I.~Staffell,
  \href{http://www.sciencedirect.com/science/article/pii/S0360544216311744}{{Long-term
  patterns of European PV output using 30 years of validated hourly reanalysis
  and satellite data}}, Energy 114 (2016) 1251 -- 1265.
\newblock \href {https://doi.org/https://doi.org/10.1016/j.energy.2016.08.060}
  {\path{doi:https://doi.org/10.1016/j.energy.2016.08.060}}.
\newline\urlprefix\url{http://www.sciencedirect.com/science/article/pii/S0360544216311744}

\bibitem{IEC61400}
{International Electrotechnical Commission},
  \href{https://webstore.iec.ch/publication/26423}{{IEC 61400-1:2019: Wind
  energy generation systems - Part 1: Design requirements}} (2019).
\newline\urlprefix\url{https://webstore.iec.ch/publication/26423}

\bibitem{Holttinen2004}
P.~Norgaard, H.~Holttinen, A multi-turbine power curve approach, 2004.

\bibitem{ENSPRESO2019}
{European Commission - Joint Research Centre},
  \href{https://data.jrc.ec.europa.eu/collection/id-00138}{{ENSPRESO - an open,
  EU-28 wide, transparent and coherent database of wind, solar and biomass
  energy potentials}} (2019).
\newline\urlprefix\url{https://data.jrc.ec.europa.eu/collection/id-00138}

\bibitem{WindDataBase}
M.~Pierrot,
  \href{https://www.thewindpower.net/store_continent_en.php?id_zone=1001}{{The
  Wind Power - Wind Energy Market Intelligence}} (2020).
\newline\urlprefix\url{https://www.thewindpower.net/store_continent_en.php?id_zone=1001}

\bibitem{BLines2018}
{Deutsche WindGuard GmbH}, \href{https://vasab.org/10564-2/}{{Capacity
  densities of European offshore wind farms}}, Tech. rep., {VASAB} (2018).
\newline\urlprefix\url{https://vasab.org/10564-2/}

\bibitem{EC_2020}
\href{https://ec.europa.eu/energy/topics/technology-and-innovation/clean-energy-competitiveness_en}{{Clean
  energy transition - technologies and innovations, accompanying the report on
  progress of clean energy competitiveness}}, Tech. rep., European Commission
  (2020).
\newline\urlprefix\url{https://ec.europa.eu/energy/topics/technology-and-innovation/clean-energy-competitiveness_en}

\bibitem{TYNDP2018}
ENTSO-E, \href{https://tyndp.entsoe.eu/maps-data}{Maps \& {D}ata: {TYNDP}2018}
  (2018).
\newline\urlprefix\url{https://tyndp.entsoe.eu/maps-data}

\bibitem{DEA2020GEN}
{Danish Energy Agency},
  \href{https://ens.dk/en/our-services/projections-and-models/technology-data/technology-data-generation-electricity-and}{{Technology
  Data for Generation of Electricity and District Heating}} (2020).
\newline\urlprefix\url{https://ens.dk/en/our-services/projections-and-models/technology-data/technology-data-generation-electricity-and}

\bibitem{ENTSOEPowerStats}
{ENTSO-E}, \href{https://www.entsoe.eu/data/power-stats/}{{Power Statistics}}
  (2021).
\newline\urlprefix\url{https://www.entsoe.eu/data/power-stats/}

\bibitem{JRC_PPDB}
{European Commission - Joint Research Centre},
  \href{https://ec.europa.eu/jrc/en/publication/joint-research-centre-power-plant-database-jrc-ppdb}{{The
  Joint Research Centre Power Plant Database (JRC-PPDB)}} (2019).
\newline\urlprefix\url{https://ec.europa.eu/jrc/en/publication/joint-research-centre-power-plant-database-jrc-ppdb}

\bibitem{JRC_HY}
{European Commission - Joint Research Centre},
  \href{https://github.com/energy-modelling-toolkit/hydro-power-database}{{JRC
  Hydro-power plants database}} (2020).
\newline\urlprefix\url{https://github.com/energy-modelling-toolkit/hydro-power-database}

\bibitem{SolarDataBase}
P.~Wolfe, \href{https://www.wiki-solar.org/data/index.html}{{Wiki Solar - The
  authority on utility-scale solar power}} (2020).
\newline\urlprefix\url{https://www.wiki-solar.org/data/index.html}

\bibitem{solarEurope2019}
A.~Beauvais, M.~Herrero~Cangas, N.~Chevillard, M.~Heisz, M.~Labordena,
  R.~Rossi,
  \href{https://www.solarpowereurope.org/eu-market-outlook-for-solar-power-2019-2023/}{{EU
  Market Outlook for Solar Power 2019-2023}}, Tech. rep., Solar Power Europe
  (2019).
\newline\urlprefix\url{https://www.solarpowereurope.org/eu-market-outlook-for-solar-power-2019-2023/}

\bibitem{dox_repo}
D.~{Radu}, A.~{Dubois}, M.~{Berger},
  \href{https://dox.uliege.be/index.php/s/u6kPFWKRG0GlDqg}{Assessing the
  economic value of renewable resource complementarity for power systems: a
  european study - dataset} (2020).
\newline\urlprefix\url{https://dox.uliege.be/index.php/s/u6kPFWKRG0GlDqg}

\bibitem{resite_ip_git}
D.~Radu, M.~Berger,
  \href{https://github.com/dcradu/resite_ip/releases/tag/v0.0.1}{resite - a
  framework for {RES} siting leveraging resource complementarity} (2021).
\newline\urlprefix\url{https://github.com/dcradu/resite_ip/releases/tag/v0.0.1}

\bibitem{replan_git}
A.~{Dubois}, D.~{Radu}, M.~{Berger},
  \href{https://github.com/montefesp/replan/releases/tag/v0.0.4}{replan - a
  framework for bulk energy systems planning and analysis} (2020).
\newline\urlprefix\url{https://github.com/montefesp/replan/releases/tag/v0.0.4}

\bibitem{Grams2017}
C.~M. Grams, R.~Beerli, S.~Pfenninger, I.~Staffell, H.~Wernli, Balancing
  europe’s wind-power output through spatial deployment informed by weather
  regimes, Nature Climate Change 7 (2017) 557–562.
\newblock \href {https://doi.org/10.1038/nclimate3338}
  {\path{doi:10.1038/nclimate3338}}.

\bibitem{Cortesi2019}
N.~Cortesi, V.~Torralba, N.~González-Reviriego, A.~Soret, F.~J. Doblas-Reyes,
  Characterization of european wind speed variability using weather regimes,
  Climate Dynamics 53 (2019) 4961–4976.
\newblock \href {https://doi.org/10.1007/s00382-019-04839-5}
  {\path{doi:10.1007/s00382-019-04839-5}}.

\bibitem{Wiser2021}
R.~Wiser, J.~Rand, J.~Seel, P.~Beiter, E.~Baker, E.~Lantz, P.~Gilman, Expert
  elicitation survey predicts 37\% to 49\% declines in wind energy costs by
  2050, Nature Energy 53 (2021) 4961–4976.
\newblock \href {https://doi.org/10.1038/s41560-021-00810-z}
  {\path{doi:10.1038/s41560-021-00810-z}}.

\bibitem{Collins2018}
S.~Collins, P.~Deane, B.~{Ó Gallachóir}, S.~Pfenninger, I.~Staffell,
  \href{https://www.sciencedirect.com/science/article/pii/S254243511830285X}{Impacts
  of inter-annual wind and solar variations on the european power system},
  Joule 2~(10) (2018) 2076--2090.
\newblock \href {https://doi.org/https://doi.org/10.1016/j.joule.2018.06.020}
  {\path{doi:https://doi.org/10.1016/j.joule.2018.06.020}}.
\newline\urlprefix\url{https://www.sciencedirect.com/science/article/pii/S254243511830285X}

\bibitem{WindPowerHub2021}
\href{https://northseawindpowerhub.eu/knowledge/towards-the-first-hub-and-spoke-project}{{Towards
  the first hub-and-spoke project}}, Tech. rep., North Sea Wind Power Hub
  Consortium (2021).
\newline\urlprefix\url{https://northseawindpowerhub.eu/knowledge/towards-the-first-hub-and-spoke-project}

\bibitem{Hartel_2017}
P.~Härtel, M.~Korpås, Aggregation methods for modelling hydropower and its
  implications for a highly decarbonised energy system in europe, Energies
  10~(11) (2017) 1841.
\newblock \href {https://doi.org/10.3390/en10111841}
  {\path{doi:10.3390/en10111841}}.

\bibitem{ENTSOETransparency}
{ENTSOE},
  \href{https://transparency.entsoe.eu/generation/r2/waterReservoirsAndHydroStoragePlants/show}{{Water
  Reservoirs and Hydro Storage Plants}} (2020).
\newline\urlprefix\url{https://transparency.entsoe.eu/generation/r2/waterReservoirsAndHydroStoragePlants/show}

\bibitem{GrandDatabase}
B.~Lehner, C.~R. Liermann, C.~Revenga, C.~Vörösmarty, B.~Fekete, P.~Crouzet,
  P.~Döll, M.~Endejan, K.~Frenken, J.~Magome, C.~Nilsson, J.~Robertson,
  R.~Rodel, N.~Sindorf, , D.~Wisser,
  \href{http://globaldamwatch.org/grand/}{High-resolution mapping of the
  world’s reservoirs and dams for sustainable river-flow management},
  Frontiers in Ecology and the Environment 9 (9) (2011) 494--502.
\newline\urlprefix\url{http://globaldamwatch.org/grand/}

\bibitem{ERA5_cds}
{ECMWF},
  \href{https://cds.climate.copernicus.eu/cdsapp#!/dataset/reanalysis-era5-single-levels?tab=overview}{{ERA5
  hourly data on single levels from 1979 to present}} (2020).
\newline\urlprefix\url{https://cds.climate.copernicus.eu/cdsapp#!/dataset/reanalysis-era5-single-levels?tab=overview}

\bibitem{eurostat_hydro}
{Eurostat},
  \href{https://ec.europa.eu/eurostat/web/products-datasets/-/nrg_ind_peh}{{Gross
  and net production of electricity and derived heat by type of plant and
  operator}} (2020).
\newline\urlprefix\url{https://ec.europa.eu/eurostat/web/products-datasets/-/nrg_ind_peh}

\bibitem{ATB2020}
NREL, \href{https://atb.nrel.gov/}{Annual technology baseline} (2020).
\newline\urlprefix\url{https://atb.nrel.gov/}

\bibitem{AEMO2020}
AEMO,
  \href{https://aemo.com.au/energy-systems/major-publications/integrated-system-plan-isp/2022-integrated-system-plan-isp/current-inputs-assumptions-and-scenarios}{2020-21
  inputs, assumptions and scenarios} (2020).
\newline\urlprefix\url{https://aemo.com.au/energy-systems/major-publications/integrated-system-plan-isp/2022-integrated-system-plan-isp/current-inputs-assumptions-and-scenarios}

\bibitem{IEA2020Costs}
{International Energy Agency},
  \href{https://www.iea.org/reports/projected-costs-of-generating-electricity-2020}{Projected
  costs of generating electricity} (2020).
\newline\urlprefix\url{https://www.iea.org/reports/projected-costs-of-generating-electricity-2020}

\bibitem{DEA2020STO}
{Danish Energy Agency},
  \href{https://ens.dk/en/our-services/projections-and-models/technology-data/technology-data-energy-storage}{Technology
  data for energy storage} (2020).
\newline\urlprefix\url{https://ens.dk/en/our-services/projections-and-models/technology-data/technology-data-energy-storage}

\bibitem{HWires}
K.~Mongird, V.~Viswanathan, P.~Balducci, J.~Alam, V.~Fodetar, V.~Koritarov,
  B.~Hadjerioua,
  \href{https://www.energy.gov/eere/water/hydrowires-publications}{Energy
  storage technology and cost characterization report} (2019).
\newline\urlprefix\url{https://www.energy.gov/eere/water/hydrowires-publications}

\bibitem{RealiseGrid}
A.~L'Abbate, G.~Migliavacca,
  \href{https://realisegrid.rse-web.it/Publications-and-results.asp}{Review of
  costs of transmission infrastructures, including cross border connections}
  (2011).
\newline\urlprefix\url{https://realisegrid.rse-web.it/Publications-and-results.asp}

\bibitem{SANDIA2013}
A.~Akhil, G.~Huff, A.~Currier, B.~Kaun, D.~Rastler, S.~B. Chen, A.~Cotte,
  D.~Bradshaw, W.~Gauntlett,
  \href{https://prod-ng.sandia.gov/techlib-noauth/access-control.cgi/2015/151002.pdf}{{DOE/EPRI
  Electricity Storage Handbook in Collaboration with NRECA}}, Tech. rep.,
  {Sandia National Laboratories} (2013).
\newline\urlprefix\url{https://prod-ng.sandia.gov/techlib-noauth/access-control.cgi/2015/151002.pdf}

\bibitem{ETSAP_TRANSMISSION}
G.~Simbolotti, G.~Tosato,
  \href{https://iea-etsap.org/E-TechDS/PDF/E12_el-t&d_KV_Apr2014_GSOK.pdf}{{Technology
  Brief E12 - Electricity Transmission and Distribution}}, Tech. rep.,
  {International Energy Agency} (2014).
\newline\urlprefix\url{https://iea-etsap.org/E-TechDS/PDF/E12_el-t&d_KV_Apr2014_GSOK.pdf}

\bibitem{TYNDP2020}
D.~McGowan, T.~Rzepczyk, C.~Sonmez, D.~Powell, M.~Fernandez, M.~Boguca,
  M.~G.~N. L.~Watine,
  \href{https://www.entsos-tyndp2020-scenarios.eu/wp-content/uploads/2019/10/TYNDP_2020_Scenario_Report_entsog-entso-e.pdf}{{TYNDP
  2020 Scenario Report}}, Tech. rep., ENTSO-E and ENTSO-G (2019).
\newline\urlprefix\url{https://www.entsos-tyndp2020-scenarios.eu/wp-content/uploads/2019/10/TYNDP_2020_Scenario_Report_entsog-entso-e.pdf}

\bibitem{IPCC2006}
D.~Gómez, J.~Watterson, A.~Branca, H.~Chia, G.~Marland, E.~Matsika,
  L.~Namayanga, O.-E. Balgis, J.-D.~K. Saka, K.~Treanton,
  \href{https://ghgprotocol.org/Third-Party-Databases/IPCC-Emissions-Factor-Database}{{IPCC
  Emissions Factor Database}}, Tech. rep., {International Panel on Climate
  Change (IPCC)} (2006).
\newline\urlprefix\url{https://ghgprotocol.org/Third-Party-Databases/IPCC-Emissions-Factor-Database}

\bibitem{EEA2020}
{European Environment Agency},
  \href{https://www.eea.europa.eu/data-and-maps/data/co2-intensity-of-electricity-generation}{{CO2
  Intensity of Electricity Generation}} (2020).
\newline\urlprefix\url{https://www.eea.europa.eu/data-and-maps/data/co2-intensity-of-electricity-generation}

\bibitem{IEA2020}
{International Energy Agency},
  \href{https://www.iea.org/data-and-statistics?country=WORLD&fuel=CO2\%20emissions&indicator=CO2\%20emissions\%20by\%20sector}{{Data
  and statistic - CO2 emissions by sector}} (2020).
\newline\urlprefix\url{https://www.iea.org/data-and-statistics?country=WORLD&fuel=CO2\%20emissions&indicator=CO2\%20emissions\%20by\%20sector}

\end{thebibliography}

\end{document}